\documentclass[12pt,preprint]{aastex}

\begin{document}

\title{The Chandra Deep Field-North Survey: \\
XVII. Evolution of magnetic activity in old late-type stars$^1$}

\author{E. D. Feigelson,\altaffilmark{2} A. E.
Hornschemeier,\altaffilmark{3,4}  G. Micela,\altaffilmark{5} F.
E. Bauer,\altaffilmark{2,3,6} D. M. Alexander,\altaffilmark{2,6}
W. N. Brandt,\altaffilmark{2} F. Favata,\altaffilmark{7}  S.
Sciortino,\altaffilmark{5} \and G. P. Garmire\altaffilmark{2}}

\altaffiltext{1}{Based in part on observations obtained with the
Hobby-Eberly Telescope, which is a joint project of the University
of Texas at Austin, the Pennsylvania State University, Stanford
University, Ludwig-Maximilians-Universität M\"unchen, and
Georg-August-Universität G\"ottingen.}

\altaffiltext{2}{Department of Astronomy \& Astrophysics, 525
Davey Laboratory, Pennsylvania State University, University Park,
PA 16802}

\altaffiltext{3}{Visiting Astronomer, Kitt Peak National Observatory,
National Optical Astronomy Observatory, which is operated by the
Association of Universities for Research in Astronomy, Inc. (AURA)
under cooperative agreement with the National Science Foundation. }

\altaffiltext{4}{Chandra Fellow, Johns Hopkins University, 3400
N.\ Charles Street, Baltimore, MD 21218}

\altaffiltext{5}{INAF - Osservatorio Astronomico di Palermo, Piazza del
Parlamento 1, 90134 Palermo, Italy}

\altaffiltext{6}{Institute of Astronomy, University of Cambridge,
Madingley Road, Cambridge CB3 0HA, United Kingdom}

\altaffiltext{7}{Astrophysics Division, Space Science Department
of ESA, ESTEC, Postbus 299, 2200 AG Noordwijk, The Netherlands}

\slugcomment{Resubmitted to ApJ, April 25 2004}

\begin{abstract}

The extremely sensitive Chandra Deep Field-North (CDF-N)
pencil-beam X-ray survey is used to identify and characterize the
X-ray emission from old high-latitude main sequence Galactic
stars. Our principal goal is to investigate the expected long-term
decay of magnetic activity of late-type stars due to the gradual
spindown of stellar rotation from a magnetized stellar wind.

Thirteen X-ray sources are associated with late-type stars; 11 of
these constitute a well-defined sample for statistical analysis.
This sample consists of 2 G, 2 K0$-$K4, and 7 M2$-$M5 stars with
median $V$-band magnitude around 19 and median distance around 300
pc. X-ray luminosities are typically $\log L_x \simeq 27$
erg~s$^{-1}$ but are substantially higher in two cases. The
combination of large-amplitude variations on timescales of hours
and plasma temperatures around $5-30$ MK indicates that the
observed X-ray emission is dominated by magnetic reconnection
flares rather than quiescent coronal emission.  These X-ray
properties are quantitatively similar to those seen in the active
contemporary Sun.

The CDF-N stellar sample is compared to simulations based on
convolution of X-ray luminosity functions (XLFs) with the known
spatial distribution of old disk stars.  The model indicates that
the CDF-N stars are the most magnetically active old disk stars. A
substantial decline in X-ray luminosities over the $1<t<11$ Gyr
age interval is required: 39 rather than 11 stars should have been
detected if the XLF does not evolve over this time interval. This
is the first demonstration that the coronal and flaring components
of stellar magnetic activity -- and presumably the interior
magnetic dynamos responsible for the reconnecting fields at the
stellar surface -- exhibit long-term decay over the age of the
Galactic disk.  The model that best fits the magnitudes, spectral
types and X-ray luminosities of the sample has $L_x \propto
t^{-2}$ erg s$^{-1}$ which is faster than the $t^{-1}$ decay rate
predicted from widely accepted rotational spindown rates and
X-ray-activity relations.

\end{abstract}


\section{Introduction \label{intro.sec}}

It is well-known from studies of young stellar clusters that magnetic
activity traced by coronal and flare X-ray emission evolves quickly
during the early stages of stellar evolution.  During the pre-main
sequence phases and upon arrival at the zero-age main sequence (ages $t
\sim 10^6-10^8$ yr), the X-ray luminosities of low-mass stars typically
lie in the range $\log L_x \sim 28 - 30$ erg s$^{-1}$ and subsequently
decay by 2 orders of magnitude over several hundred million years
\citep{Micela85, Gudel97, Feigelson99, Micela02}.

Late-type stellar magnetic activity is regulated principally by
rotation in main sequence stars \citep{Pallavicini81, Noyes84,
Baliunas95}, and its decay with stellar age is attributed to
rotational spin-down. While the rotational evolution during the
earlier phases of stellar evolution is complex due to magnetic
coupling to the circumstellar disk and changing internal
structure, the angular momentum behavior is thought to be simpler
once the star settles on the main sequence. The only process
thought to be important is rotational braking from mass loss where
ionized wind particles gain large specific angular momentum as
they travel outward along spiral-shaped magnetic field lines that
corotate with the star \citep{Schatzman62}.

Model calculations indicate that surface rotational velocities
should decline as $v_{rot} \propto t^{\alpha}$ with $-0.75 <
\alpha < -0.38$ depending on whether the magnetic field geometry
is closer to a dipole or radial configuration \citep{Kawaler88}
which agrees with empirical findings that $\alpha \simeq -0.5$
\citep{Skumanich72}. Models considering possible differences in
the pre-main sequence disk-locking time, solid body $vs.$
differential rotation in the interior, and the onset of magnetic
saturation give decay laws over the range $-0.8 < \alpha < -0.2$
for $1-5$ Gyr old solar-mass stars \citep{Krishnamurthi97}.
\citet{Barnes03} argues that the angular momentum loss from the
magnetized wind efficiently couples to the entire stellar
interior.  X-ray coronal and flaring luminosities are
well-established tracers of surface magnetic activity, and it is
empirically found that $L_x \propto v_{rot}^\beta$ on the main
sequence where $\beta \simeq 2$ \citep{Pallavicini81,
Pizzolato03}. Combining this with the calculated spin-down
relation, the X-ray emission from a typical older main sequence
star is expected to decline as
\begin{equation}
L_x \propto t^{\alpha \times \beta} \propto t^{-1}
\end{equation}
if stars spin-down as $\alpha \simeq -1/2$ and X-ray luminosity
scales with rotation as $\beta \simeq 2$.

There have been relatively few empirical measures of this expected
decline in magnetic activity in older late-type stars, and most of
these involve chromospheric activity. The survey of CaII H and K
emission of older solar-mass stars by \citet{Soderblom91} found an
activity decay of $R^\prime_{HK} \propto t^{-2/3}$ out to $t
\simeq 10$ Gyr. The low-level activity of two slowly rotating
analogs of the ancient Sun, $\beta$ Hyi ($t = 6.7$ Gyr) and 16 Cyg
A ($t \sim 9$ Gyr), has been studied in detail \citep{Guinan03}.
However, other chromospheric studies suggest that the decay of
activity may flatten to a constant level after $t \sim 3-5$ Gyr
and/or that a subpopulation of chromospherically active old stars,
possibly from coalesced binaries, is present \citep{Smith98,
RochaPinto02}.

With respect to X-ray emission, even the existence of activity
decay between 1 and 10 Gyr has not yet been confidently
established. Several dM halo stars observed with the $ROSAT$
satellite are systematically fainter ($25.5 < \log L_x < 26.7$ erg
s$^{-1}$) and have cooler ($T \simeq 2-6$ MK) coronal plasmas than
in old-disk dM stars ($27.0 < \log L_x < 28.5$ erg s$^{-1}$, $T >
6$ MK; Micela, Pye \& Sciortino 1997). Old-disk stars, in
contrast, are found to have $ROSAT$ X-ray luminosities and flares
similar to the general disk population. Using the $ROSAT$ All-Sky
Survey dataset and another sample of 24 solar-mass stars with
isochronal ages, \citet{Micela02a} found no clear decay law over
the range $1<t<10$ Gyr.  $ROSAT$ All-Sky Survey observations of a
large sample of $Hipparcos$ F stars showed, contrary to
expectations, that X-ray luminosity is correlated (rather than
anticorrelated) with increased age inferred from kinematics
\citep{Suchkov03}. Most studies of the evolution of X-ray activity
indicators use only a rough grouping of stars into logarithmic age
intervals (e.g., $<0.01$, $0.1-1$, $1-10$ Gyr) and thus are unable
to address the activity decay expected at later ages.

This issue can be addressed by statistical analyses of deep
flux-limited X-ray surveys of random locations rather than shallow
all-sky surveys or pointings at known stars. While the stellar
X-ray source counts in shallow wide-field X-ray surveys are
dominated by the X-ray bright young disk stars, models convolving
measured XLFs with Galactic structure predict that older disk
stars should dominate the source counts at fainter flux levels
\citep{Favata92, Micela93, Sciortino95, Guillout96, Kuntz01}. Even
the simple measurement of the total number of stellar sources at
faint flux levels can be a sensitive indicator of magnetic
activity evolution.  At fluxes around $10^{-16}$ erg cm$^{-2}$
s$^{-1}$ at high Galactic latitudes, stellar soft X-ray source
counts are dominated by older ($3<t<10$ Gyr) M dwarfs. If the
X-ray decay index $\alpha \times \beta \simeq -1$ over $1<t<10$
Gyr for these low mass stars, the stellar source counts would be
reduced several fold compared to a constant-activity model (see
Figure 1 of Kuntz et al.). A handful of individual late-type stars
has been found in past deep high-latitude pencil-beam X-ray
surveys \citep[e.g.,][]{Griffiths83, Lehmann01, Stern02}, but
these samples have been too small or incomplete for statistical
analysis.

Here we address the question of X-ray evolution of older stars
with the extremely sensitive $Chandra$ Deep Field-North (CDF-N)
pencil-beam survey towards $(l,b) = (126, 55)$ which probes the
high-galactic latitude X-ray sky to an unprecedented level of $3
\times 10^{-17}$ cm$^{-2}$ erg s$^{-1}$ in the soft $0.5-2$ keV
band \citep{Alexander03}. A critically important feature is that
optical spectroscopy has been performed for all of the X-ray
sources with $V < 22.5$ \citep{Barger02}. While the great majority
of the CDF-N sources are extragalactic
\citep[e.g.,][]{Hornschemeier00, Brandt01, Bauer02, Alexander03},
we find that $\simeq 3$\% are low-mass stars in the outer disk of
our Galaxy. Although the sample of 13 stars (11 in our statistical
sample) is small, it is the largest well-defined sample of its
type and sufficient to provide new constraints on the evolution of
X-ray properties over the age of the Galactic disk. We find that
the overall stellar XLF must decline over this age range, perhaps
even faster than expected with $\alpha \times \beta \simeq -2$
rather than $\simeq -1$ (equation 1), but that hot plasma
temperatures and high-amplitude magnetic reconnection flares
persists for long times in at least some stars.

Sections $2-3$ below present the stellar X-ray sample, its X-ray
properties, and the associated optical counterparts. Qualitative
implications are discussed in \S 4, and quantitative modelling is
presented in \S 5.  Concluding comments appear in \S 6.

\section{$Chandra$ observations and analysis \label{xray.sec}}

\subsection{The CDF-N survey and resulting stellar sample
\label{ACISobs.sec}}

The X-ray data utilized here are part of the 2~Ms CDF-N survey obtained
with the Advanced CCD Imaging Spectrometer (ACIS) detector on board the
{\it Chandra X-ray Observatory}.  The satellite and detector are
described by \citet{Weisskopf02}.  Extragalactic results from the CDF-N
project are described in other papers in this series. We briefly
summarize the observations and data reduction here.  The full
observation log, areal exposure, reduction details, and source list are
given by \citet{Alexander03}.

The data were collected during 20 separate observations with the
ACIS imaging array between November 1999 and February 2002 with a
total field of view of $\approx 448$ arcmin$^2$.  The effective
exposure time ranges up to 1.95~Ms; about half of the field has an
effective exposure time exceeding 1~Ms. The data were corrected
for charge transfer inefficiency, cleaned of spurious events, and
limited to the $0.5-8$ keV band. The background level is small
($<1$ ct pixel$^{-1}$) even in the region of highest exposure.
Unresolved sources were located in the field using a wavelet-based
algorithm. This located 503 unresolved X-ray sources over the
entire \hbox{CDF-N} field, about 370 of which were previously
reported in the 1~Ms CDF-N sample by \citet{Brandt01}. The field
was accurately aligned to the FK5 reference frame via radio source
counterparts. The median uncertainty of individual source
positions is $\simeq 0.3$\arcsec, although faint off-axis sources
can have individual offsets as large as $\simeq$1\arcsec. The
statistical sample of stellar sources discussed below is based on
the first 1 Ms source list \citep{Brandt01} which had been
optically characterized at the time of this study, although the
full 2 Ms dataset was available for X-ray analysis.

It is critical for our modelling of the stellar source population
to know the sky coverage as a function of X-ray flux limit. We
derived this sensitivity curve for the 1~Ms exposure following the
2~Ms survey methods described by \citet{Alexander03}.  The
differential sensitivity curve was binned into values listed in
Table~\ref{sens.tab}.  The precision of this curve is not
well-established as it is based on a complex wavelet source
detection procedure applied to a field with an inhomogeneous
exposure map.  We estimate that the sensitivity values could be
inaccurate by as much as $\pm 30$\%.

We have extracted our sample of X-ray selected stars as follows.
Optical counterparts for CDF-N sources were found down to a limit
of $V \simeq 27.3$, and spectra for 85 of the 1~Ms sources with
$V<22.5$ were obtained or reported by
\citet{Barger02}.\footnote{The spectroscopic coverage is 98\%
complete.  Two sources with $V<22.5$, CXOHDFN J123706.7+622549 and
J123828.4+620903, were missed.}  Seventy-three of these spectra
showed emission lines from active galactic nuclei and/or were
associated with morphologically extended galaxies with significant
redshifts.  The remaining 11 optical spectra show photospheric
absorption lines at zero redshift characteristic of normal
Galactic stars. Some additional spectra of counterparts from the
CDF-N 2~Ms sample \citep{Alexander03} were also obtained, from
which two more star-like spectra were found.

\begin{figure}
\centering
\includegraphics[width=1.0\textwidth]{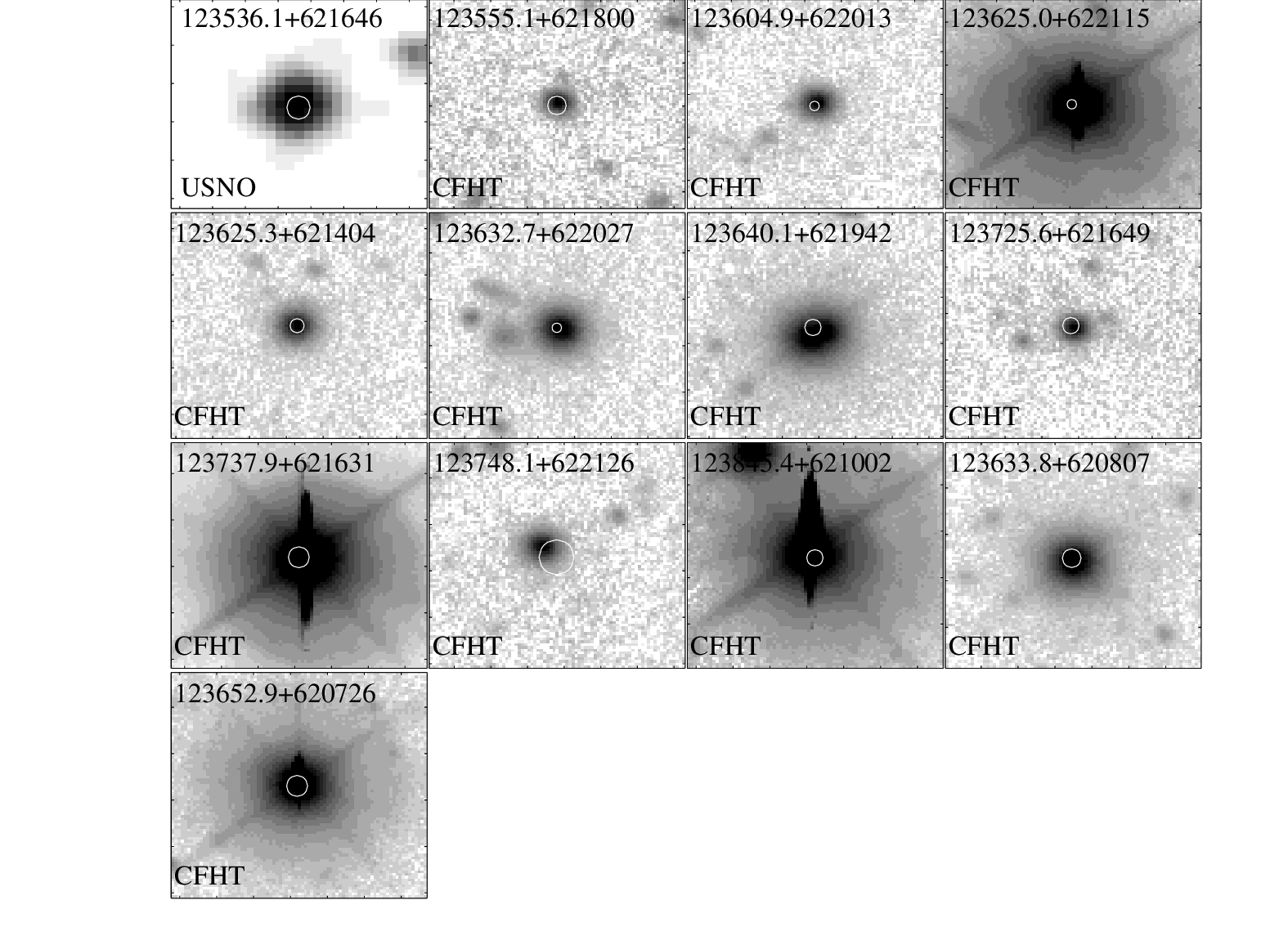}
\caption{Chandra ACIS source positions (white circles) superposed on
$I$-band images (kindly provided by G.\ Wilson), except for the first
panel where the Palomar $R$-band sky survey image is shown.  Each
panel is $25\arcsec \times 25\arcsec$ with North above and East to the
right.  The error circle radii are the X-ray positional errors given in
Table \ref{xray.tab}. \label{offsets.fig}}
\end{figure}

We thus emerge with a well-defined or `statistical' sample of 11
stellar sources from the CDF-N 1~Ms survey, and two additional
stellar sources from the 2~Ms survey. Finding charts are shown in
Figure \ref{offsets.fig}. These sources are listed in Table
\ref{xray.tab} with several X-ray properties extracted from the
tables of \citet{Alexander03}: the celestial position; counts in
the full band ($FB$, $0.5-8$ keV), soft band ($SB$, $0.5-2$ keV)
and hard band ($HB$, $2-8$ keV); the band ratio $HB/SB$; the
$0.5-8$ keV count rate; and the effective exposure $t_{eff}$ at
that location in the image. The first column ``B01 \#'' gives the
source identification number from \citet{Brandt01}.  Note that,
unlike most extragalactic sources, there were often no hard-band
events detected.

None of our stars are included in the $V$- and $I$-band sample of
stellar-like objects found in $\simeq 25$ arcmin$^2$ covering the
Hubble Deep Field-North and its flanking fields reported by
\citet{Mendez98}. Most of our stars are brighter than the stars
considered in that study.

\subsection{X-ray spectra and fluxes}

We analyze here the X-ray spectra using the response matrices
appropriate for CTI-corrected ACIS data \citep{Townsley02} and the
XSPEC spectral fitting package \citep{Arnaud96}.  We adopted
thermal plasma models of \citet{Mewe91}.  No spectrum required
significant interstellar absorption, consistent with the low
Galactic column density, $N_H = (1.6 \pm 0.4) \times
10^{20}$~cm$^{-2}$ \citep{Stark92},  in the CDF-N direction.
Two-temperature models were used when the least-squares solution
to one-temperature models gave unsatisfactory $\chi^2$
goodness-of-fit values.

\begin{figure}
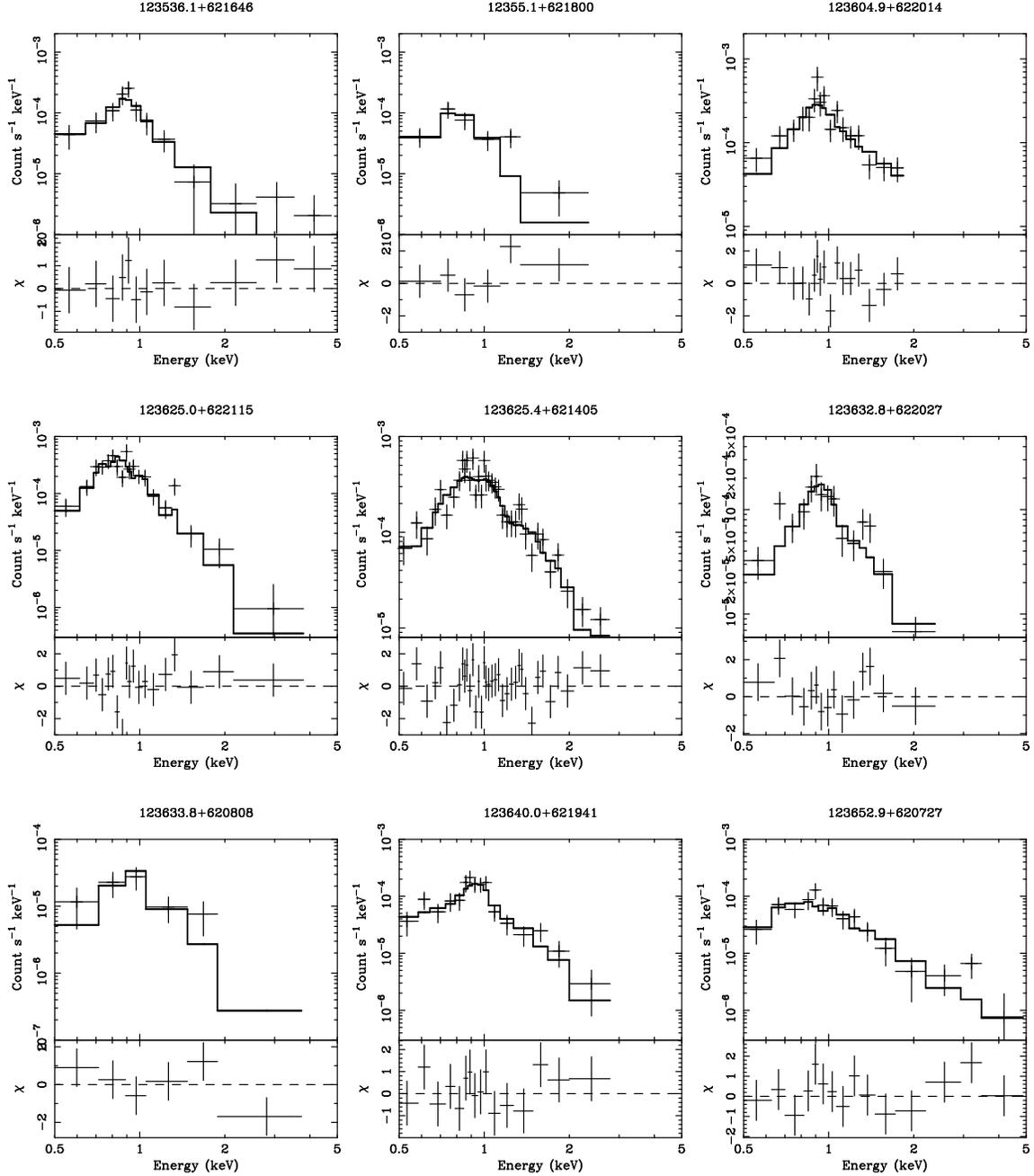

 \begin{minipage}[t]{1.0\textwidth}
  \centering
    \includegraphics[angle=-90., width=0.30\textwidth]{f2a.eps}
    \includegraphics[angle=-90., width=0.30\textwidth]{f2b.eps}
    \includegraphics[angle=-90., width=0.30\textwidth]{f2c.eps}
 \end{minipage} \\ [0.2in]
 \begin{minipage}[t]{1.0\textwidth}
  \centering
    \includegraphics[angle=-90., width=0.30\textwidth]{f2d.eps}
    \includegraphics[angle=-90., width=0.30\textwidth]{f2e.eps}
    \includegraphics[angle=-90., width=0.30\textwidth]{f2f.eps}
 \end{minipage} \\ [0.2in]
 \begin{minipage}[t]{1.0\textwidth}
  \centering
    \includegraphics[angle=-90., width=0.30\textwidth]{f2g.eps}
    \includegraphics[angle=-90., width=0.30\textwidth]{f2h.eps}
    \includegraphics[angle=-90., width=0.30\textwidth]{f2i.eps}
\caption{ACIS spectra of the 13 CDF-N stars with best-fit plasma
models (Table \ref{xray_prop.tab}) superposed.  The lower part of
each panel shows the deviation of the data from the model in units
of $\sigma$. \label{acis_spec.fig}}
 \end{minipage}
\end{figure}

The ACIS spectra for the 13 stellar sources are shown with the
best-fit spectral models in Figure \ref{acis_spec.fig}. Associated
spectral parameters are listed in Table \ref{xray_prop.tab}:
plasma energies for the one- or two-temperature model; the reduced
$\chi^2$ value and associated degrees of freedom of the fit; and
the $SB$ and $FB$ fluxes obtained by integrating the model.  The
large uncertainties in the temperatures show that a wide range of
spectral models can be fit to the data.  In a few cases,
additional spectral components which are not statistically
significant may be present (see the notes to Table
\ref{xray_prop.tab}).

\subsection{X-ray variability } \label{xvar.sec}

As the arrival time for each X-ray event is recorded to within $\sim 3$
sec precision, light curves can be constructed. Long-term lightcurves
of the background-corrected count rates seen in each CDF-N observation
(not shown) show that most of the sources show factors $\geq 2$
variability on timescales of months.  Such behavior is expected from
any type of stellar magnetic activity.

It is of greater interest that examination of the photon arrival
times within each exposure showed a number of likely short-term
flares. Figure \ref{xray_flares.fig} and Table \ref{xray_prop.tab}
show cases where the statistical significance of a 1-sample
Kolmogorov-Smirnov test of the hypothesis of constant emission has
a probability $P_{KS} < 0.001$.  We have verified the reality of
these flares using the Bayesian Block analysis code developed by
\citet{Scargle98}. Altogether there are 9 flares in 7 sources
ranging in amplitude from factors of $\approx$3--10. Most of the
flares have short durations of $\approx$1--3 hours, although one
source, CXOHDFN J123625.4+621405, shows a longer-term decay of
$\sim 1$ day (Figure~\ref{xray_flares.fig}).

\section{Optical and near-infrared data \label{optir.sec}}

\subsection{Spectroscopy \label{optspec.sec}}

From the extensive optical spectroscopy campaign conducted for
CDF-N X-ray sources, we show in Figure~\ref{opt_spec.fig} optical
spectra for all 13 stars.  Twelve of the spectra presented here
were obtained with the Marcario Low Resolution Spectrograph (LRS)
\citep{Hill98} of the Hobby-Eberly Telescope (HET) 9 meter
telescope \citep{Ramsey98}.  The queue-scheduled HET observations
occurred on multiple evenings between 2000 February 7 and 2002
March 23.  A 2\farcs0 slit and 300~line~mm$^{-1}$ grism/GG385
blocking filter produced spectra from 4400--9000~\AA\ at 17~\AA\
resolution; data redward of  7700~\AA\ are suspect because of
possible second-order contamination.  The exposure time per source
ranged from 2--20~min.  The seeing was typically 2\farcs5 (FWHM).
Wavelength calibration was performed using HgCdZn and Ne lamps,
and relative flux calibration was based on spectrophotometric
standards.

The remaining star, CXOHDFN~J123652.9+620726, was observed with
the Ritchey Cretien Spectrograph on the KPNO Mayall 4~m telescope
on 2002 May 15.  The LB1A CCD detector, BL400 158~line~mm$^{-1}$
grating, OG530 blocking filter and 1\farcs8 slit were used.  This
configuration produced spectra from 5300--10000~\AA\ at 14.5~\AA\
resolution. Wavelength calibration was performed using HeNeAr
lamps, and relative flux calibration was performed using
spectrophotometric standards.

\subsection{Photometry \label{phot.sec}}

Photometric observations are presented for the stellar sample in
Table \ref{phot.tab}.  For objects fainter than $\sim 18$ mag,
$UBVRz^\prime$ magnitudes are obtained from \citet{Barger02} which
was based on the 1~Ms X-ray CDF-N catalog of
\citet{Brandt01}.\footnote{As all of the stellar sources discussed
here fall outside the Hubble Deep Field-North and its Flanking
Fields \citep{Williams96}, deep Hubble Space Telescope data were
not available when this analysis was made. We also do not use the
$UBR$ measurements of \citet{Liu99}, as these were affected by bad
weather, nor the $U_nGRK_s$ photometry of \citet{Hogg00}.}  For
brighter objects where the \citet{Barger02} observations are
saturated, we use the $V$ and $I$-band photometry of
\citet{Barger99} in the inner 9\farcm0$\times$9\farcm0 of the
field, and our own measurements of the Canada-France-Hawaii
Telescope UH8K camera $V$ and $I$-band images obtained by
\citet{Wilson03} which cover $\approx90$\% of CDF-N field. The
photometry on the UH8K $V$ and $I$-band images was performed using
{\sc sextractor} \citep{Bertin96} with the ``Best" magnitude
criteria, a $2\sigma$ detection threshold, and a 25-pixel sigma
Gaussian wavelet.

The accuracies of these optical photometric measurements, whose
primary purpose was to identify the extragalactic populations in
this survey field, are not sufficiently high or well-established
for detailed stellar characterization.  For example, a
cross-calibration of the UH8K and \citet{Barger99} $VI$ values
shows a scatter around 0.3~mag.  We thus cannot reliably search
for luminosity class or metallicity effects from the optical
photometry.

In the near-infrared $JHK$ bands, photometry is obtained for most
of our sample from the 2 Micron All-Sky Survey
\citep[2MASS;][]{Cutri00}. In two faint cases, magnitudes are
converted from the $HK^{\prime}$ measurements of \citet{Barger02}
using the conversion $K \simeq HK^{\prime} - 0.3$
\citep{Barger99}.

\subsection{Reliability of the X-ray/stellar associations}

Table \ref{phot.tab} gives the positional offsets between the
$Chandra$ and stellar sources; most of the latter positions are
from the 2MASS survey with typical $\pm 0.2$\arcsec\/
uncertainties with respect to the {\it Hipparcos} reference frame.
One source, \#341 (CXOHDFN J123748.1+622126), has an unusually
large offset (1.8\arcsec) which is probably acceptable because the
source lies $\sim 10$\arcmin\/ off-axis in the $Chandra$ field
where the point spread function is broad.  We conclude that the
identification of these X-ray sources with Galactic stars is
highly reliable.

\subsection{Derived quantities}

Spectroscopic classifications were obtained by visual comparison
of the spectra in Figure~\ref{opt_spec.fig} to low-resolution
spectral atlases \citep{Jacoby84, Kirkpatrick91, Silva92}.  The
results are reported in Table \ref{deriv.tab}.  We also attempted
spectral classification from the photometry in Table
\ref{phot.tab};  most classifications were consistent with the
spectroscopic classifications, but classification accuracy was
lower and inconsistencies were found.  Most of these problems are
probably due to inaccurate photometric measurements, although the
possibility of binarity cannot be excluded.

If we assume that all stars lie on the main sequence and that
negligible interstellar absorption is present, distances can be readily
estimated.  For the majority of the sample that are M stars, we use the
tabulations of \citet{Bessell91} who gives $BVRIJHK$ colors, absolute
magnitudes $M_I$ and bolometric corrections $BC_I$ for the $I$ band
where the bulk of the flux emerges.  Because the 2MASS $K$-band
magnitudes are more accurate than the available $I$-band measurements,
we base our distance calculation on $K$ with a classification-dependent
$(I-K)$ correction.  Distances for the G and K stars were estimated
from $K$ magnitudes with classification-dependent $(V-K)$ corrections
obtained from \citet{Bessell90} and absolute $M_V$ luminosities given
in Table 15.3.1 of \citet{Cox00}.

These distances are then used to convert the time-averaged soft-band
X-ray fluxes from Table \ref{xray_prop.tab} to soft X-ray luminosities
$\log L_{SB}$. Results are tabulated in the final column of
Table~\ref{deriv.tab}.  From systematic uncertainties in spectral
typing, photometry and X-ray spectral fitting, we estimate that
distances are uncertain by $\pm 30$\% and log X-ray luminosities by
$\pm 0.3$ (total range).  From the X-ray light curves, we know that the
instantaneous luminosity of a given star may differ by several fold
from the listed time-averaged value.

We have also calculated the X-ray/near-infrared flux ratio for each
star, using the 2MASS filter parameters of \cite{Cohen03}. The relation
is as follows:
\begin{equation}
\log{ {f_{\rm X}}\over{f_K} } = \log{f_{\rm X}} + K/2.5 + 6.95
\end{equation}

\section{Results and comparison with the Sun \label{res_Sun.sec}}

We restrict our interpretation to the well-defined and reliable
sample of 11 stars from the 1~Ms CDF-N survey with a peak
sensitivity around $3 \times 10^{-17}$ erg cm$^{-2}$ s$^{-1}$ in
the soft X-ray band. Here we have two G stars, two K0$-$K4 stars,
and seven M2$-$M5 stars. Distances range from 50$-$460 pc with a
median around 300 pc (Table \ref{deriv.tab}).  Visual magnitudes
range from $V=14-22$ with a median around 19 (Table
\ref{phot.tab}). Soft band X-ray luminosities range from $\log
L_{SB} = 26$ to nearly 30 erg s$^{-1}$ with a median around
$L_{SB} \simeq 5 \times 10^{27}$ erg s$^{-1}$.  Most stars show
plasma components with temperatures $T \sim 2-30$ MK, and the
majority show flare events with amplitudes around $\log
L_{SB,peak} \sim 27.5-28.5$ erg s$^{-1}$ (Table
\ref{xray_prop.tab}).

Nine of the 11 stars have X-ray luminosities in the range $26.1 <
\log L_{SB} < 28.1$ erg s$^{-1}$ which is essentially the range
exhibited by the contemporary Sun through its 22-year cycle
\citep{Peres00}.  The two most distant stars have higher
luminosities , \#341 with $\log L_{SB} = 28.8$ and \#370 with 29.8
erg s$^{-1}$.  This is consistent with the expected bias between
sensitivity and distance for a flux-limited survey.  The observed
plasma temperatures ranging from $1-30$ MK are also consistent
with the range seen on the Sun from its quiet, maximum, and
flaring states.

If one considers the 11 stars as a homogeneous ensemble, then the
flare amplitudes and frequencies of old-disk stars in the CDF-N
field can be estimated.  We detect nine X-ray flares (\S
\ref{xvar.sec}) with characteristic peak luminosities $\log
L_{SB,peak} \simeq 28$ erg s$^{-1}$, and one approaching 30 erg
s$^{-1}$, in a total exposure of $\simeq 22$ Ms.  This gives an
estimated flare frequency of one flare every $\simeq 2.4$ Ms above
$\simeq 10^{28}$ erg s$^{-1}$. This is similar to the rate of
high-X-ray-luminosity flares on the Sun: from several years of
$GOES$ satellite data, \citet{Sammis00} find $\simeq 20$ solar
flares with $28.0 \leq \log L_{FB}(peak) \leq 29.0$ erg s$^{-1}$
in the $1.6-12$ keV band (close to the $Chandra$ full band) or
about 1 flare every $\simeq 1.5$ Ms.

We thus find that the flare amplitudes, frequencies and
temperatures of the CDF-N old disk stars (most of which are lower
mass K- and M-stars) are remarkably similar to the strongest
flares seen in the contemporary Sun. This result differs from that
reported in $ROSAT$ studies of nearby old disk and halo M dwarfs.
These show weaker emission around $L_{SB} \sim 2 \times
10^{26}$~erg~s$^{-1}$ and unusually soft emission \citep{Hawley94,
Micela97}.  This is not necessarily a contradiction. The
characteristics of the nearby sample, containing well-known
high-proper motion stars such as Kapteyn's star and Barnard's
star, can be roughly viewed as a volume-limited sample
representative of average older stars independent of their X-ray
luminosities.  Our CDF-N sample, in contrast, is a flux-limited
X-ray survey capturing only stars at the top of the stellar XLF.
Also, our older thin-disk stars may be generally younger than the
nearby sample which contains stars kinematically classified as
thick disk and halo stars.  We thus infer that, while our sample
resembles the contemporary Sun, the average old-disk star probably
has weaker X-ray emission and magnetic activity.

\section{The decay of stellar X-ray emission \label{quan_model.sec}}

In order to make inferences about the evolution of magnetic
activity from these findings, it would be valuable to determine
individual ages for the stars in our sample.  Unfortunately, this
is difficult to achieve.  The stars are not sufficiently high
above the Galactic plane to be likely members of the thick disk or
halo which are known to be ancient ($t \simeq 10-13$ Gyr)
\citep{Majewski93}. While the maximum age of stars in the thin
disk is $11 \pm 1$ Gyr \citep{Binney00}, we expect a strong age
(and thus mass) gradient with height $z$ above the Galactic plane
because of the gradual stochastic scattering of stars to orbits
reaching several hundred parsecs above the plane. The scattering
is also strongly dependent on mass, producing the long-known link
between spectral type and exponential scale height.  For a typical
partitioning of star counts into thin disk, thick disk and halo
components, the mass-averaged stellar age rises from $t<0.5$ Gyr
for $z < 100$ pc to $<t> \simeq 2$ Gyr at $z = 150$ pc, $<t>
\simeq 4$ Gyr at $z = 200$ pc, and $5-11$ Gyr for $z > 300$ pc
\citep[][H.\ Rocha-Pinto, 2003 private
communication]{RochaPinto00, Robin03}. From these broad
distributions, we conclude that the CDF-N stars most likely have
ages in the range $3<t<11$ Gyr, but that individual ages can not
be estimated.\footnote{Star \#157 may be an exception; at $d
\simeq 50$ pc ($z \simeq 40$ pc), its location suggests it may be
a young disk star with age $t < 0.5$ Gyr. However, it has the
lowest X-ray luminosity of the sample and thus does not exhibit
the enhanced magnetic activity of younger stars. We thus suspect
it is an old disk star like the others in the CDF-N sample.}

In the absence of individual stellar age estimates permitting a
direct plot of X-ray emission $vs.$ age, we compare the observed
distribution of CDF-N stars to those predicted by simulated
populations assuming different magnetic activity decay laws, $L_x
\propto t^{\alpha \times \beta}$, as presented in \S
\ref{intro.sec}. This approach is feasible only because the CDF-N
survey is complete both in its X-ray source population and in the
identification of its stellar counterparts, within well-defined
flux limits (\S \ref{xray.sec}-\ref{optir.sec}). We implement this
strategy using the {\it XCOUNT} model developed by
\citet{Favata92} and \citet{Micela93} which convolves evolving
stellar XLFs with the spatial distribution of stars in the Galaxy
along a chosen line of sight. {\it XCOUNT} gives predictions of
the magnitudes, spectral types, distances and X-ray luminosites of
stars detected in a flux-limited X-ray observation at a chosen
Galactic location $(l,b)$. Except for the time dependency of the
XLF of interest, there are no free parameters in the model.  Even
though our sample is small, comparing the CDF-N results with model
predictions can be powerful: any discrepancy in the X-ray flux,
spectral type or magnitude distribution can falsify a model
calculation and lead to insights into the evolution of stellar
X-ray emission in the $1<t<11$ Gyr age range.

\subsection{Description of the {\it XCOUNT} model
\label{xcount_desc.sec}}

The version of {\it XCOUNT} utilized here includes only a thin
exponential disk following the Galactic model of \citet{Bahcall80}
as revised by \citet{Gould97}.  The scale heights are also a
function of stellar age: for GKM-type stars, we set $h = 120$,
200, 400 pc for ages $0.01-0.1$, $0.1-1$, and $>1$ Gyr,
respectively.  This age dependence has little effect on optical
star counts but is very important for X-ray counts at low Galactic
latitudes because young stars are much brighter than older stars
\citep{Micela93}.  Giants are omitted from the {\it XCOUNT}
calculation here because their typical X-ray emissivity is $-6 <
\log L_x/L_{bol} < -5$ \citep{Pizzolato00} so that only giants
with $V \leq 13$ could be detected in the CDF-N survey, none of
which are present in the field.  Close binary systems, which may
be X-ray luminous either by enhanced magnetic activity or
accretion, are also omitted due to their rarity.  The {\it XCOUNT}
model includes a spatial model for interstellar matter,
responsible for absorption both in X-ray and visual bands, but
this has no importance for our high-latitude study here.  The
Besan\c{c}on model of \citet{Robin03} gives a similar number and
distribution of stars within $d \simeq 600$ pc, but also includes
a large number of thick-disk and halo stars extending out many
kiloparsecs.\footnote{This is based on a simulation of the
Besan\c{c}on model produced interactively at
\url{www.obs-besancon.fr/modele/model2003.html}.} Their omission
from the {\it XCOUNT} simulations should not impact the simulation
here because such distant stars would be detected in the CDF-N
survey only if their X-ray luminosities greatly exceeded the
magnetic activity saturation limit $\log L_t/L_{bol} \sim -3$ of
late-type stars \citep[e.g.][]{Vilhu87}.

In the standard settings of the {\it XCOUNT} model, the XLFs are
set as follows: the youngest stars with ages $0.01-0.1$ Gyr are
assigned the $ROSAT$ XLF of Pleiades stars \citep{Micela96};
somewhat older stars with ages $0.1-1$ Gyr are assigned the
$ROSAT$ XLF of the Hyades \citep{Stern95}; and stars older than 1
Gyr are assigned the XLF of $Einstein$ studies of nearby old-disk
stars \citep{Schmitt85, Maggio87, Barbera93}.  This standard model
thus has no magnetic activity decay during the $1<t<11$ Gyr age
range of interest here.  X-ray decay laws were implemented in 1
Gyr bins starting with the Hyades XLF and decreasing the
individual X-ray luminosities by the factor $t^{\alpha \times
\beta}$.  Stellar ages $t$ were randomly selected from a
multi-exponential disk assuming scale heights $h=150$ pc for $1-2$
Gyr stars, $h=200$ pc for $2-5$ Gyr stars, and $h=300$ pc for
$5-10$ Gyr stars \citep{RochaPinto00}.  Trials were made for two
decay laws: $\alpha \times \beta = -1$ as expected from past
studies (\S \ref{intro.sec}) and a faster decay law with $\alpha
\times \beta = -2$.

The {\it XCOUNT} model was then applied to give a simulated
stellar population with distances and X-ray luminosities in the
direction of the CDF-N survey, $(l,b)=(126,55)$. We then exclude
stars with X-ray fluxes below the CDF-N sensitivity limit given in
Table \ref{sens.tab} and with $V$-band magnitudes below the
optical spectroscopic limit of $V=22.5$ (\S \ref{optspec.sec}).
These two criteria effectively reproduce the sensitivity limits of
our study.

\subsection{Model results \label{xcount_results.sec}}

The results of the {\it XCOUNT} simulation are shown in Figure
\ref{model.fig}.  The top histograms (thin solid lines) show the
underlying disk population in the CDF-N field of view with
$V<22.5$. There are about 91 stars, mostly early M-type which, in
the standard model with no XLF evolution for ages $t>1$ Gyr, have
mostly low X-ray luminosities in the range $26 < \log L_x < 28$
erg s$^{-1}$. The next histogram (dot-dashed line) shows that 39
of the 90 stars would be detected with the CDF-N X-ray sensitivity
curve (Table \ref{sens.tab}) in the standard no-evolution model.
The detected stars would include most of the G- and K-type stars
but less than half of the M stars.  As expected in a flux-limited
survey, the X-ray detected sample would be biased toward the
high-$L_x$ end of the underlying XLF.

The prediction of 39 detections from the standard {\it XCOUNT}
model with no late XLF evolution is clearly much higher than the
observed X-ray selected population of 11 stars.  To determine
whether this is a significant difference, we attempt to evaluate
the sources of uncertainty in the observed numbers. On the
observational side, there is a systematic uncertainty in the CDF-N
areal coverage $vs.$ sensitivity curve (\S \ref{ACISobs.sec}) and
a statistical uncertainty in the 11-star sample.  The areal
coverage has a plausible range of $\pm 30$\% arising primarily
from the ill-determined sensitivity of the wavelet source
detection across the ACIS field.  The 90\% Poisson confidence
interval for 11 observed stars is $7-16$ stars.  On the modelling
side, there are systematic uncertainties in the Galactic structure
model and adopted XLFs, as well as statistical $\sqrt{N}$
uncertainties in the {\it XCOUNT} simulation.  We estimate a 90\%
range of $\pm 30$\% for these uncertainties.  A rough combination
of these statistical and systematic sources of uncertainties gives
an estimated range of $80-100$\%, or an approximate 90\%
confidence interval of $5-22$ stars.

We thus conclude that the prediction of 39 stars from the model
without $t>1$ Gyr magnetic activity evolution is confidently
falsified. The dashed histograms in Figure \ref{model.fig} show
that the rapid-evolution model where $\alpha \times \beta = -2$ in
the relation $L_x \propto t^{\alpha \times \beta}$  matches the
observed distributions very well.  It predicts 14 X-ray detected
stars, consistent with the observed 11 stars.  The predicted
distributions of apparent magnitude, spectral type and X-ray
luminosity (and, implicitly, stellar distances) are almost
identical to the observed distributions.  The $\alpha \times \beta
= -1$ model predictions lie between the no-evolution $\alpha
\times \beta =0$ and rapid-evolution $\alpha \times \beta = -2$
models.  It overpredicts the X-ray sample size and gives a poor
match to the spectral type and X-ray luminosity distributions, but
it is not definitely inconsistent with the observations given the
estimated uncertainties.

\section{Discussion}

\subsection{The detected stellar population}

We present the first quantitative study of stellar sources in a
deep X-ray pencil-beam high-latitude survey designed to
investigate the evolution of coronal and flare magnetic activity
in older late-type stars.  It is based on the extremely deep
$Chandra$ image of the $Chandra$ Deep Field-North field with
optical photometry and spectroscopy of all X-ray sources down to
$V = 22.5$.  From the 1 Ms exposure, we find 11 stars which
constitute our statistical sample. Two additional X-ray stars are
identified in the 2 Ms exposure.

The 11 stars consist of two G, two K and seven M stars with
photometric distances ranging from $50-510$ pc, assuming they lie
on the main sequence. At these distances, luminosities fall in the
range $26.1 < \log L_x < 29.8$ erg s$^{-1}$ ($0.5-8$ keV band) and
are consistent with those seen in the integrated field population
of nearby disk stars \citep{Schmitt97}. The luminosities,
temperatures and flaring amplitudes and frequencies of the CDF-N
stars are also remarkably similar to those of the contemporary Sun
(\S \ref{res_Sun.sec}).

Comparing the observed stars to the expected Galactic population
along the CDF-N line of sight, we infer that the stars are mostly
dwarfs of the old-disk population with characteristic ages in the
$3<t<11$ Gyr range (\S \ref{quan_model.sec}). The CDF-N
observation is insufficiently sensitive to detect the thick-disk
or halo populations, and the field is too small to contain more
than $0-1$ young ($t<1$ Gyr) thin disk stars.  We thus provide
here an empirical demonstration that stellar coronae and flares --
and presumably the interior magnetic dynamos responsible for the
reconnecting fields at the stellar surface -- persist over the age
of the Galactic disk.  However, the detected sample comprises only
the most X-ray-luminous $\simeq 10$\% within the old-disk
population. The typical old-disk stars probably have X-ray
luminosities considerably below the median value $L_{SB} \simeq 5
\times 10^{27}$ erg s$^{-1}$ of our sample.

\subsection{Magnetic activity decay on the main sequence}

Quantitative modelling of the CDF-N sample, including the spectral
types, distances and X-ray luminosities, clearly requires
substantial decline in X-ray luminosities over the $1<t<11$ Gyr
age interval.  This is the first empirical demonstration of the
long-term decay of the coronal and flaring components of magnetic
activity.  An excellent fit to the sample is obtained with $\log
L_x \propto t^{\alpha \times \beta}$ erg s$^{-1}$ with an $\alpha
\times \beta = -2$ decay law (\S \ref{xcount_results.sec}, Figure
\ref{model.fig}).  However, due to the small sample and systematic
uncertainties, we cannot confidently exclude the $\alpha \times
\beta= -1$ decay law which is expected from a simple application
of $L_x \propto v_{rot}^{\beta}$ with $\beta = 2$ X-ray
activity-rotation relation and a $v_{rot} \propto t^{\alpha}$
rotation-age relation with $\alpha = -1/2$ (\S \ref{intro.sec}).

If the $\alpha \times \beta= -2$ result is confirmed to be correct
by future studies (\S \ref{future.sec}), we will face the
challenge of explaining why the decay of magnetic activity is much
faster than the expected value $\alpha \times \beta = -1$.  One
class of explanations is that the rotational spindown for older
disk stars is more rapid than the $\alpha = -1/2$ rate expected
from constant magnetized stellar winds \citep{Kawaler88}.  It is
possible that the coupling efficiency between outer and inner
layers of the stars weakens with age, or that it is mass-dependent
\citep[see discussion in][]{Barnes03}.  Alternatively, the Sun's
wind strength may be lower than in these magnetically active old
disk stars, or coronal mass ejections may contribute significantly
to stellar angular deceleration.

The alternative class of explanations is that the magnetic
activity-rotation relation steepens with age so that $\beta \simeq
4$.  Perhaps magnetic field generation at the tachocline becomes
more efficient as the stars age and their convective zone thicken.
Coronal densities and volumes may scale differently with rotation
as stars age \citep[see discussion in][]{Ivanova03}.

\subsection{Flaring in old stars}

The extremely long CDF-N exposures give us unprecedented ability
to study the temporal behavior and spectra of X-ray emission from
older stars. We see high-amplitude flares with $L_{x,peak} \sim
10^{27}$ erg s$^{-1}$ (and in one case, $L_{x,peak} \sim 10^{30}$
erg s$^{-1}$) in half of the 11 CDF-N stars (\S \ref{xvar.sec}).
This is a good indication that magnetic dynamos, likely solar-type
$\alpha-\Omega$ dynamos arising from differential rotation at the
radiative-convective zone interface, are responsible for the X-ray
emission.  Our results do not support a view where only a
quiescent basal chromosphere \citep[e.g.,][]{Buchholz98} persists
in all older main sequence stars.  At least in the 10\% most
active stars in the $3<t<11$ Gyr range, solar-type flaring is
common.

The presence of stellar flares in old disk stars also provides new
empirical information relevant to the enigmatic star-to-star
variations in lithium abundances in old-disk and halo stars
\citep{Spite82}. While several models for the lithium
inhomogeneities have been considered, one possibility is that the
lithium is produced by spallation in magnetic reconnection flares
\citep{Ryter70, Montes98}.

\section{Avenues for future research \label{future.sec}}

High-resolution spectroscopic study of the 11 X-ray-selected CDF-N
stars would yield important insights into the magnetic activity of
older stars.  Measurement of surface rotational velocities
$v$sin$i$ would allow direct measurement of $\beta$, the X-ray
activity-rotation relation. Measurement of low metallicities
and/or high radial velocities might reveal that some stars are
members of the ancient thick-disk or halo populations. For the G
and K stars with significant $B$-band flux, measurement of the
chromospheric indicator $\log R^{\prime}_{HK}$ would permit
comparison of chromospheric and coronal/flare magnetic activity.
Measurement of a strong Li $\lambda$6707 line might test lithium
spallation models.  Obtaining sufficiently high-signal and
high-resolution spectra on such faint stars is a significant
observational challenge.

A clear limitation of the present study is its small 11-star
sample. One consequence of this small sample is our confluence of
dwarf G, K and M stars to estimate a single magnetic activity
decay rate, while it is quite possible that the magnetic evolution
is mass dependent.  If the sample of X-ray-selected old-disk stars
could be increased several fold, then considerably more powerful
constraints on the evolution of magnetic activity would emerge.
The demographic approach based on star counts in deep X-ray
pencil-beam surveys, can be extended in three ways.  First, the
CDF-N itself can be investigated at a deeper flux level using the
full 2 Ms observation, lower signal-to-noise sources, and stacking
techniques to obtain sample-averaged X-ray emission of previously
identified stars in the field. The $Chandra$ Deep Field-South
offers a similar opportunity.\footnote{The $Chandra$ Deep
Field-South catalog of \citet{Giacconi02} has several sources with
soft spectra and relatively bright optical counterparts which are
good candidates for older Galactic stars. The brightest cases are
CXOCDFS J033249.8-275455 \citep[K0 V, $V=16.16$, $U-B=0.29$,
$B-V=0.50$, $V-R=0.56$, $V-I=1.01$, $J=14.62$,
$K=21.03$;][]{Groenewegen02} and CXOCDFS J033242.1-275704
\citep[$V=14.00$, $B-V=2.00$;][]{Wolf01}.} Proper motions derived
from multi-epoch optical observations of the GOODS survey regions
may also give valuable constraints on the kinematics of X-ray
selected high-latitude stars.

Second, while our count rates indicate that $Chandra$ observations
at high Galactic latitude with exposure $\leq 0.3$ Ms will give a
low return on stellar sources, such shorter exposures at
intermediate latitudes which intersect greater column densities of
old disk stars may be successful.  For example, two Galactic
dwarfs are found among 153 sources in the 0.18~Ms Lynx field at $b
= 39^\circ$ \citep{Stern02}. Large-scale spectroscopic studies of
serendipitous X-ray sources, such as the $Chandra$ Multiwavelength
Project \citep{Green04} and $XMM-Newton$ Serendipitous Sky Survey
\citep{Watson03}, could produce quite rich stellar samples.
Intermediate- and high-latitude surveys with the EPIC imaging CCD
on the $XMM-Newton$ satellite are particularly promising, given
the detector's larger field of view compared to $Chandra$. One
coronal star has been found among the first 27 $XMM-Newton$
sources studied in the medium-sensitivity high-latitude $AXIS$
survey \citep{Barcons02}. We thus look forward to both $Chandra$
and $XMM-Newton$ studies to substantially extend the present study
of the late evolution of stellar magnetic activity.

{\bf Acknowledgments} ~~ The first three authors contributed
equally to this study.  Pat Broos and Leisa Townsley (Penn State)
played critical roles in developing data analysis techniques,
Gillian Wilson (Hawaii) kindly provided access to unpublished
optical images, and Helio Rocha-Pinto and Steve Majewski
(Virginia) provided helpful advice on Galactic structure.  The
anonymous referee gave a thoughtful and useful commentary on the
manuscript.  The Marcario Low Resolution Spectrograph is named for
Mike Marcario of High Lonesome Optics who fabricated several
optics for the instrument but died before its completion. The LRS
is a joint project of the Hobby-Eberly Telescope partnership (the
University of Texas at Austin, the Pennsylvania State University,
Stanford University, Ludwig-Maximilians-Universit\"at M\"unchen,
and Georg-August-Universit\"at G\"ottingen) and the Instituto de
Astronom\'ia de la Universidad Nacional Autonoma de Mexico. The
HET is named in honor of its principal benefactors, William P.
Hobby and Robert E. Eberly.  We gratefully acknowledge the
financial support of NASA grant NAS~8-38252 (GPG, PI), NASA GSRP
grant NGT5-50247 and $Chandra$ Fellowship PF2-30021 (AEH), CXC
grant GO2-3187A and NSF CAREER Award AST 9983783 (FEB, DMA, WNB).
GM and SS acknowledge financial support from ASI contracts and
MIUR PRIN grants.

\newpage

\clearpage
\newpage

\clearpage
\newpage

\begin{figure}
 \begin{minipage}[t]{1.0\textwidth}
  \centering
    \includegraphics[angle=-90., width=0.30\textwidth]{f2j.eps}
    \includegraphics[angle=-90., width=0.30\textwidth]{f2k.eps}
    \includegraphics[angle=-90., width=0.30\textwidth]{f2l.eps}
 \end{minipage} \\ [0.2in]
 \begin{minipage}[t]{1.0\textwidth}
  \centering
    \includegraphics[angle=-90., width=0.30\textwidth]{f2m.eps}
 \end{minipage}
\end{figure}

\clearpage
\newpage

\begin{figure}
  \centering
 \begin{minipage}[t]{1.0\textwidth}
  \centering
  \includegraphics[width=0.3\textwidth]{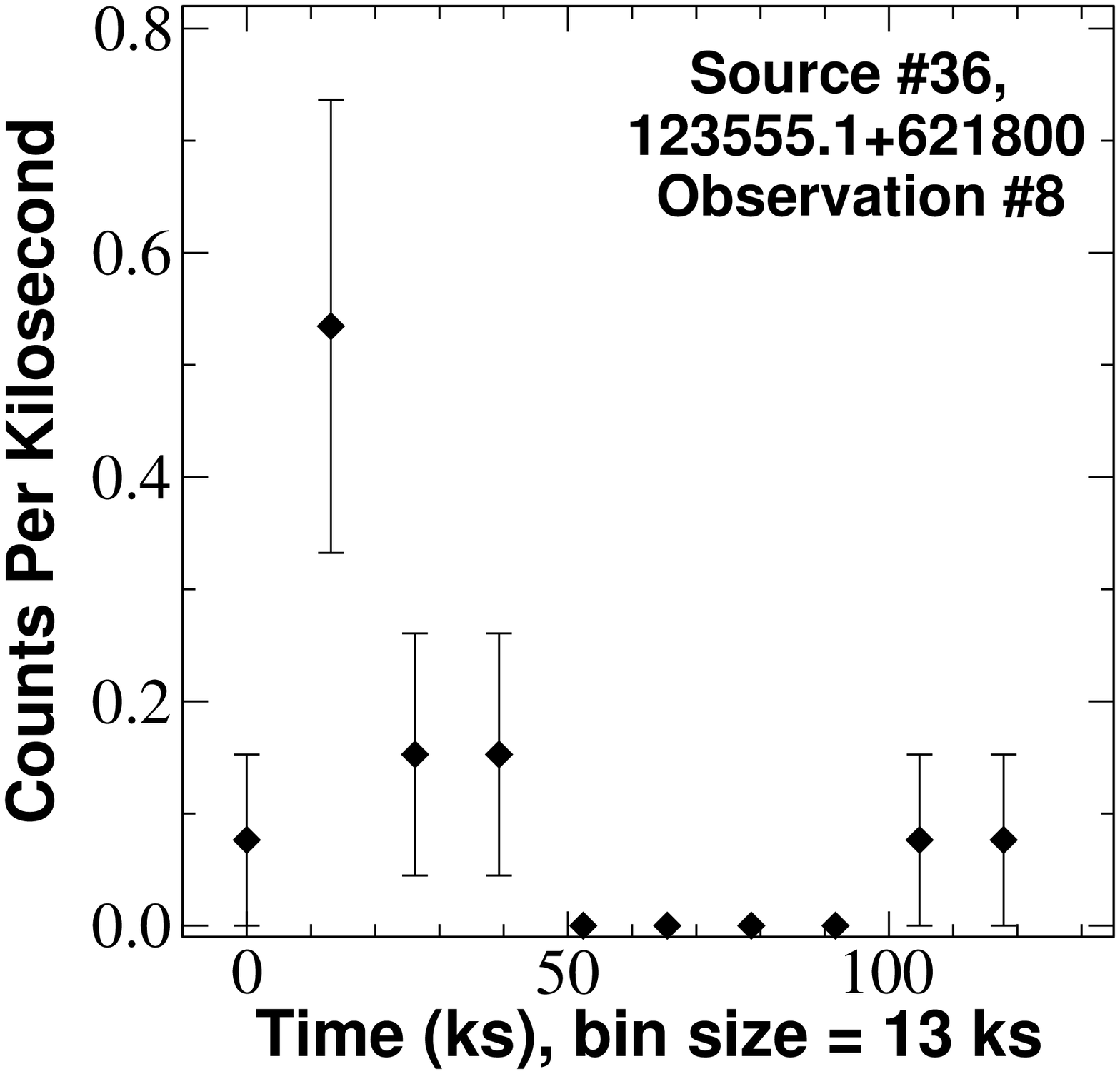}
  \includegraphics[width=0.3\textwidth]{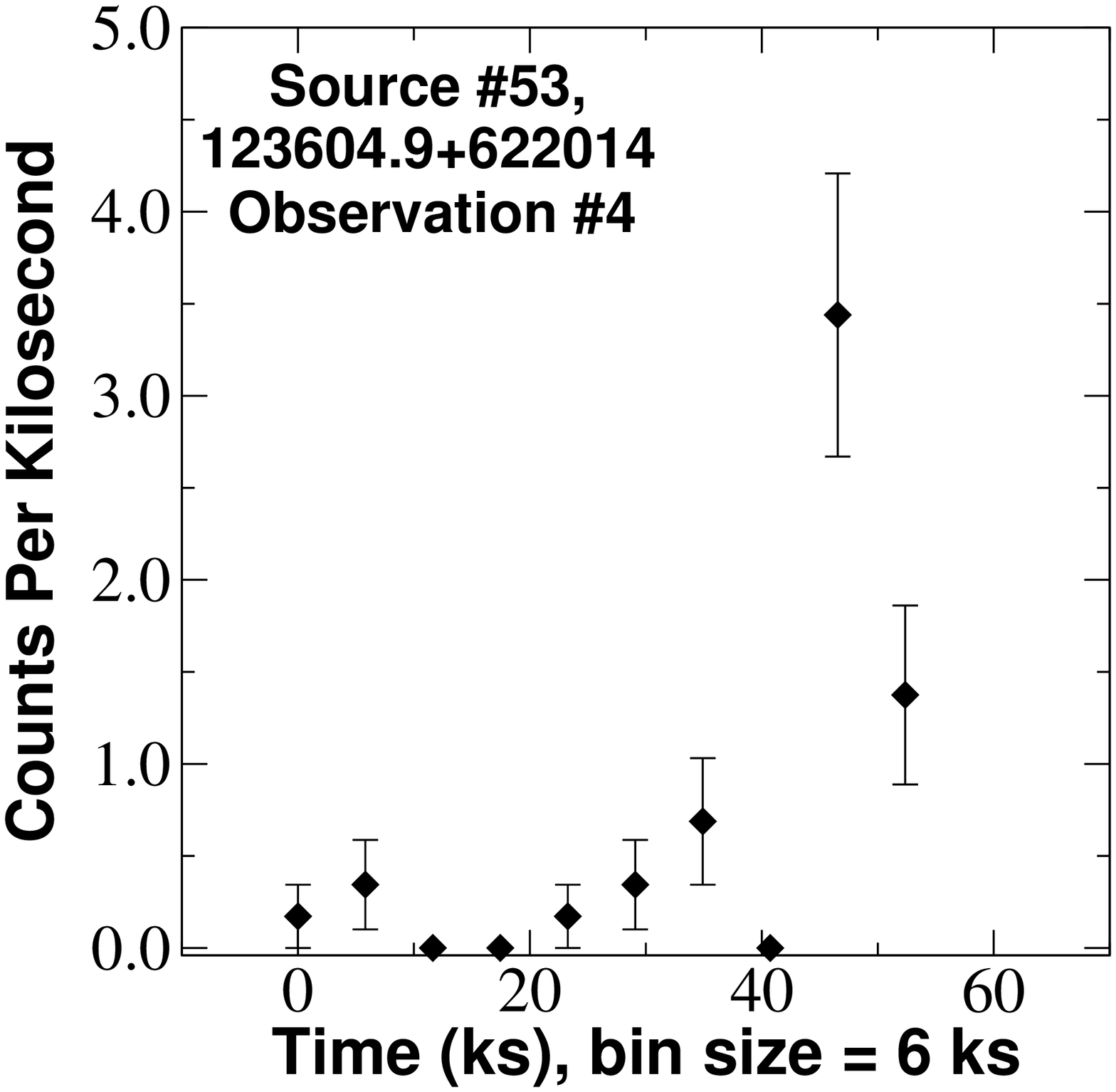}
  \includegraphics[width=0.3\textwidth]{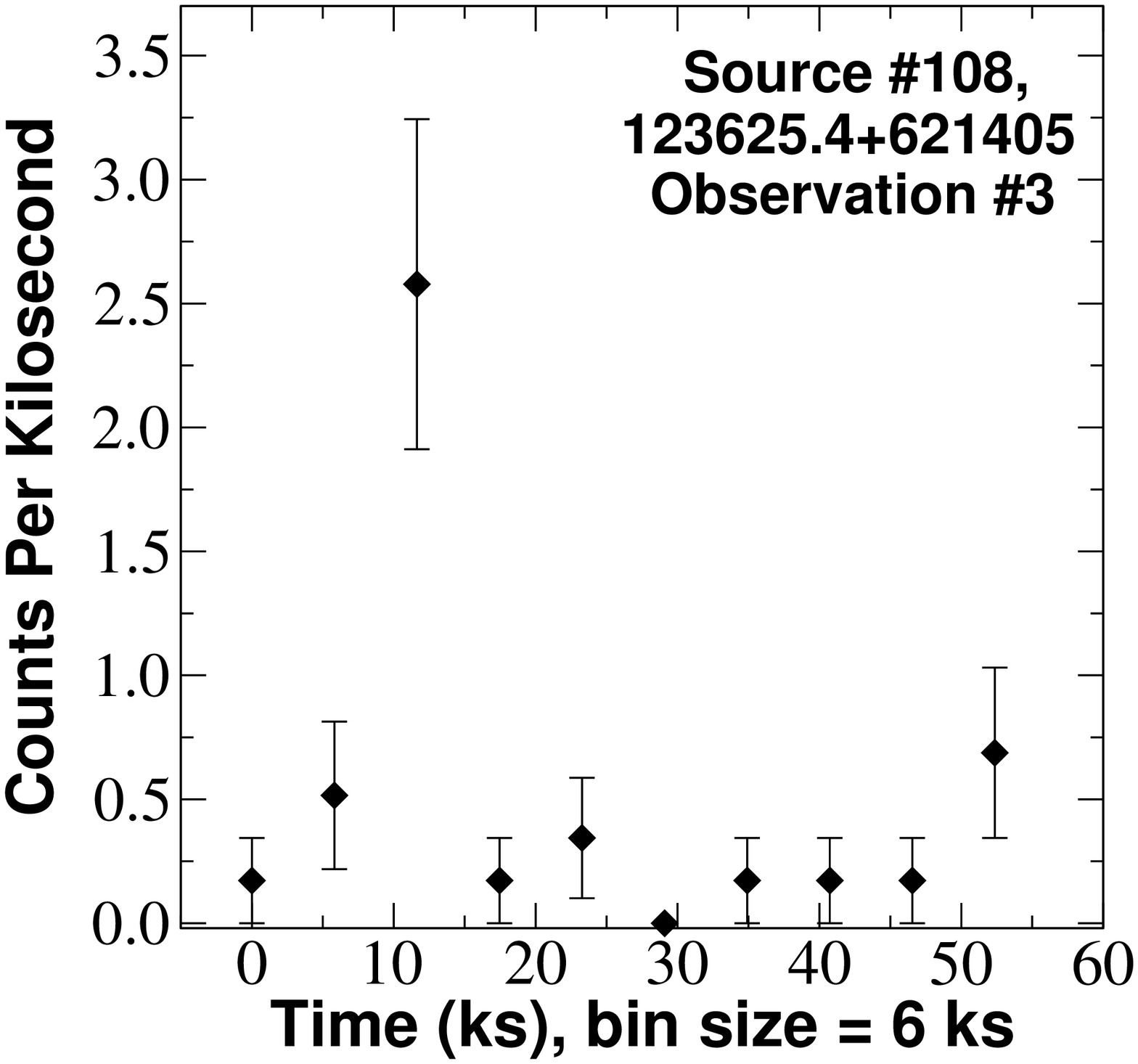}
 \end{minipage} \\ [0.2in]
 \begin{minipage}[t]{1.0\textwidth}
  \centering
  \includegraphics[width=0.3\textwidth]{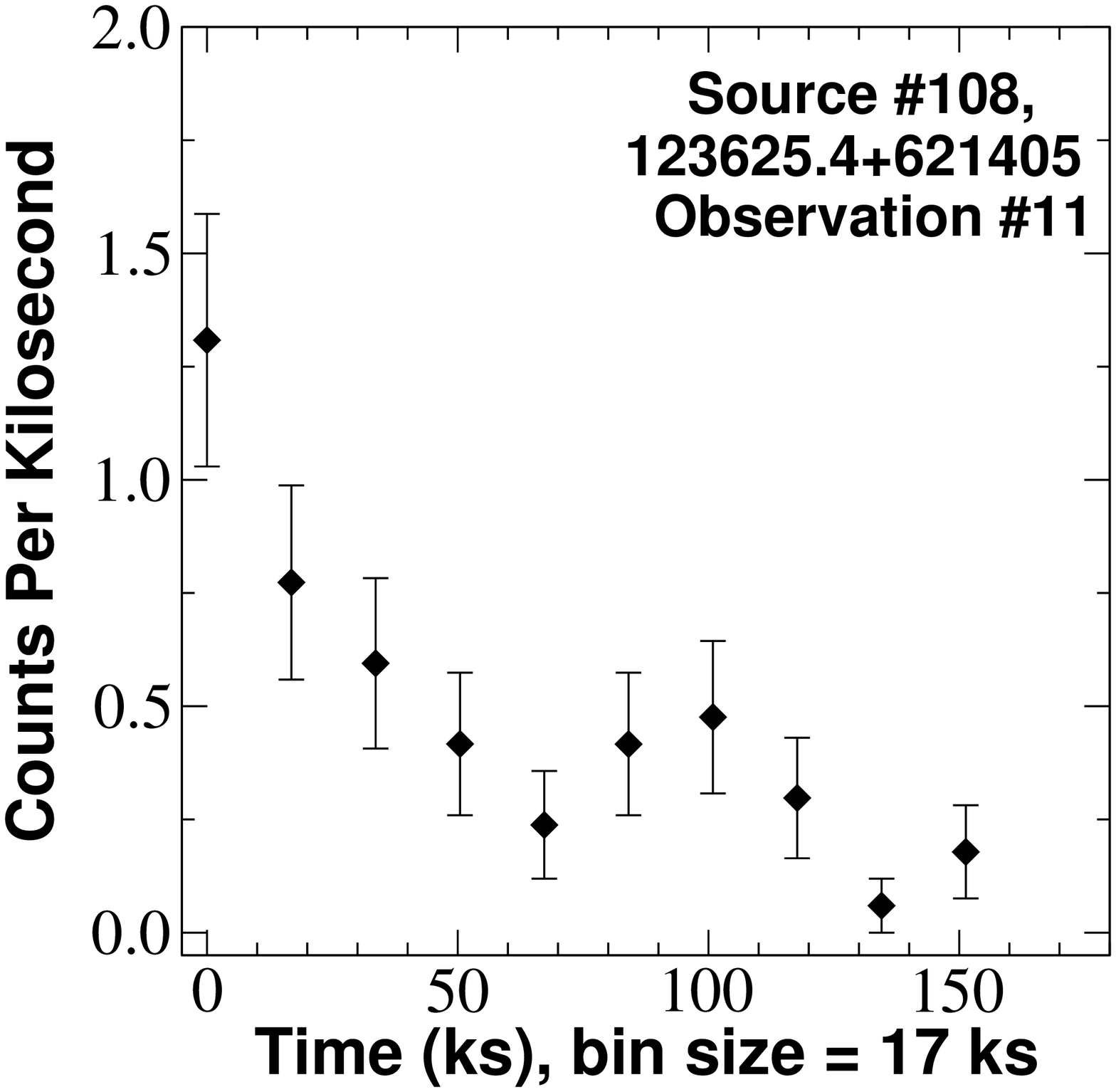}
  \includegraphics[width=0.3\textwidth]{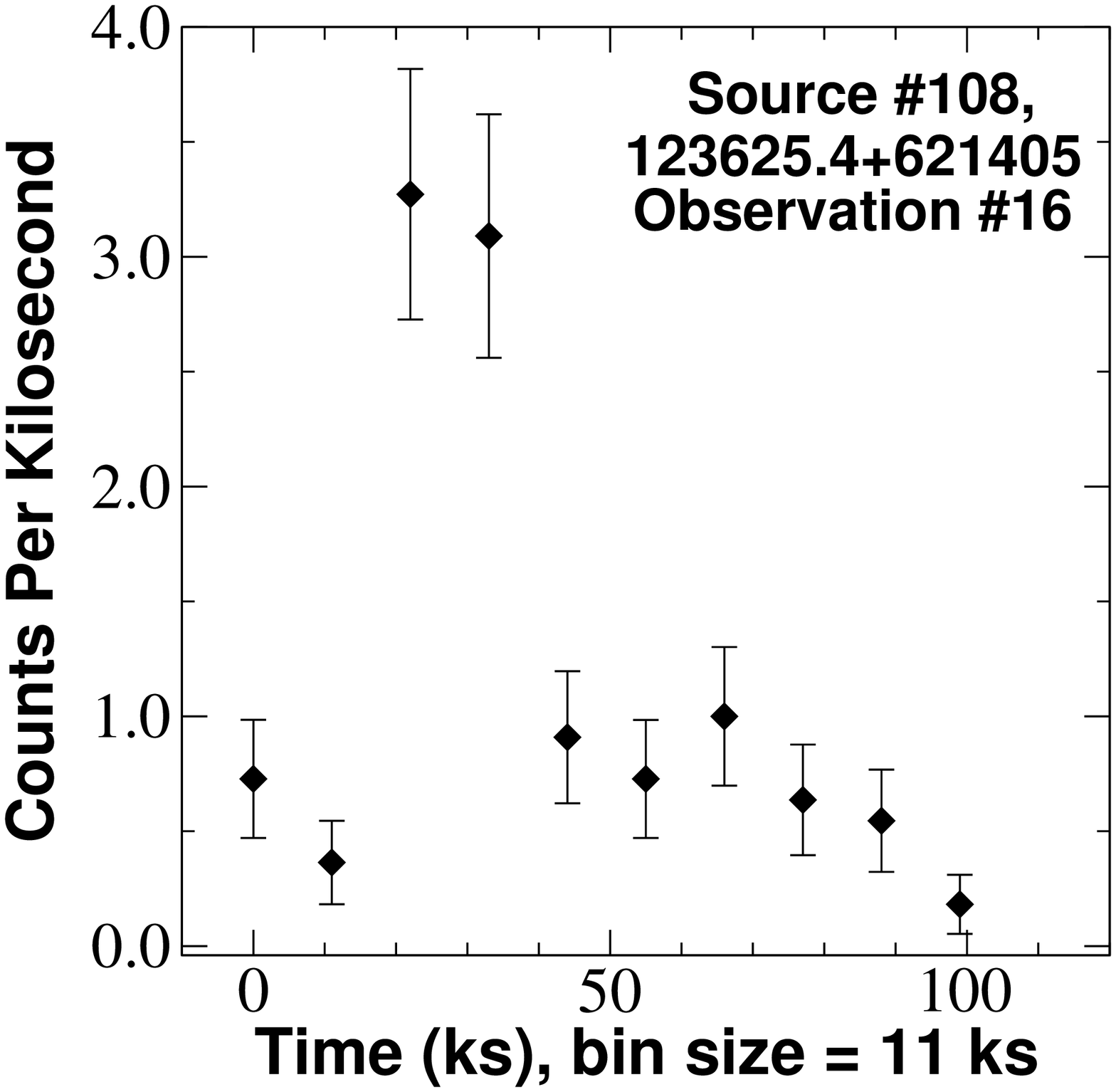}
  \includegraphics[width=0.3\textwidth]{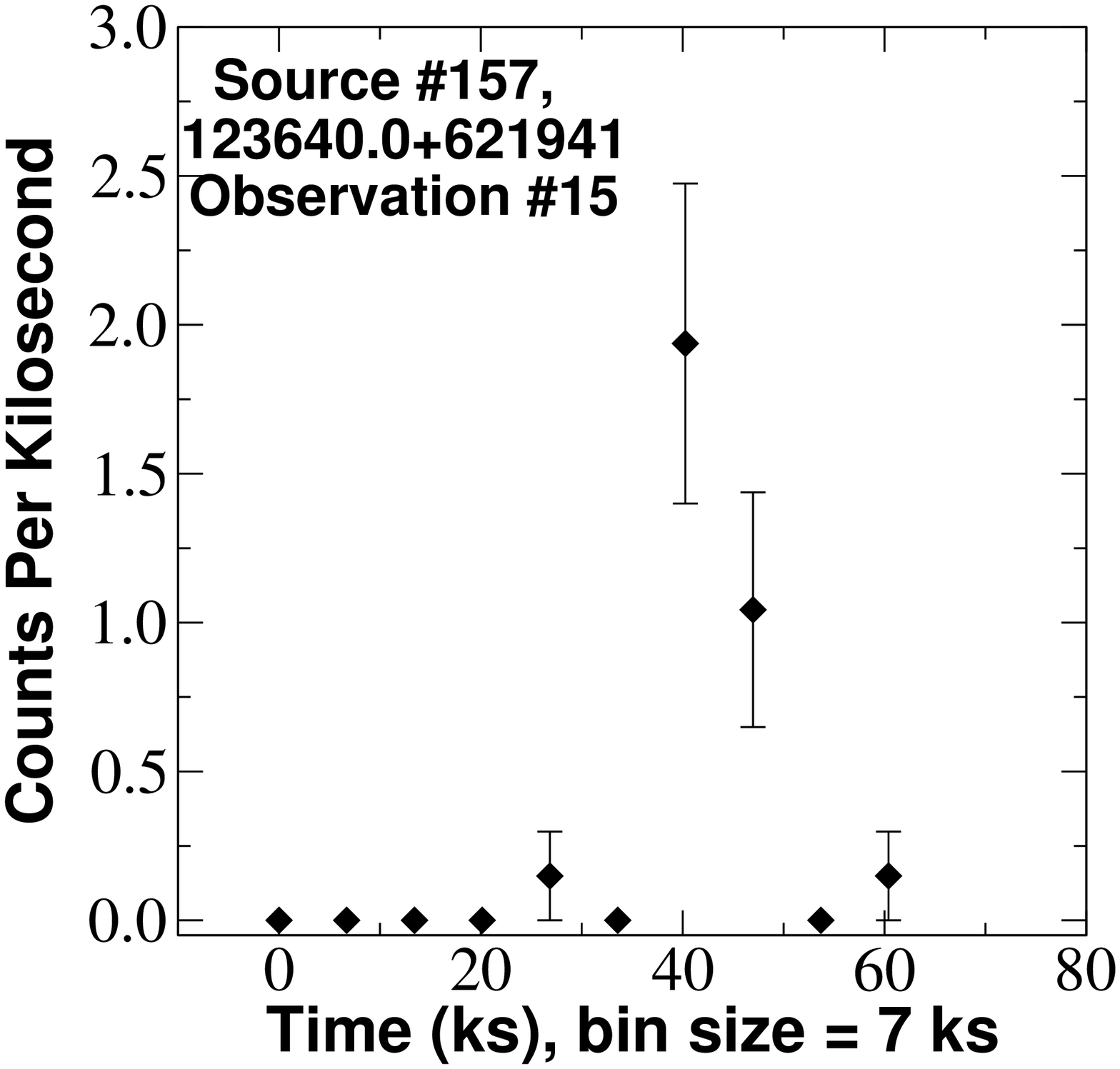}
 \end{minipage} \\ [0.2in]
 \begin{minipage}[t]{1.0\textwidth}
 \centering
  \includegraphics[width=0.3\textwidth]{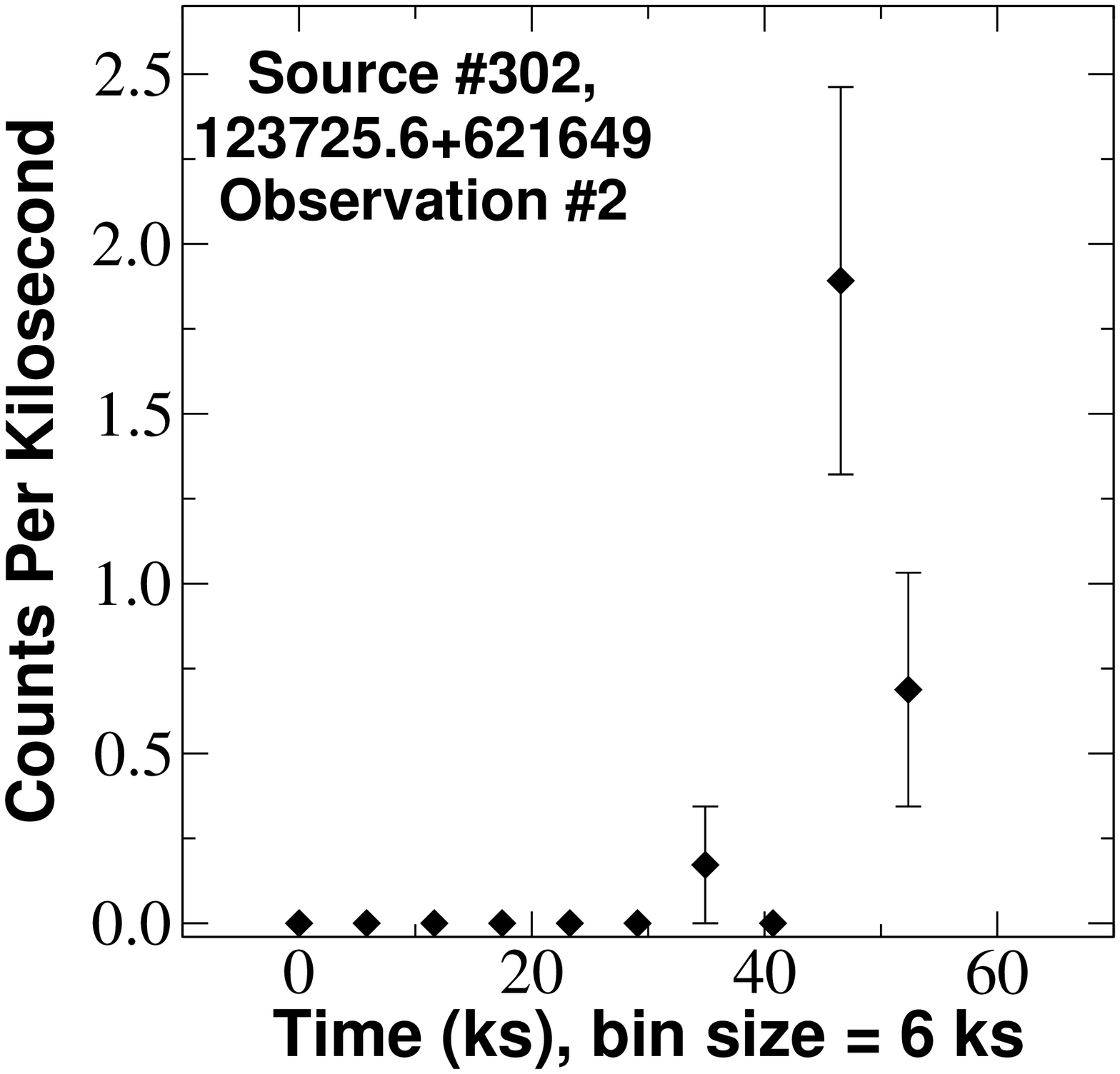}
  \includegraphics[width=0.3\textwidth]{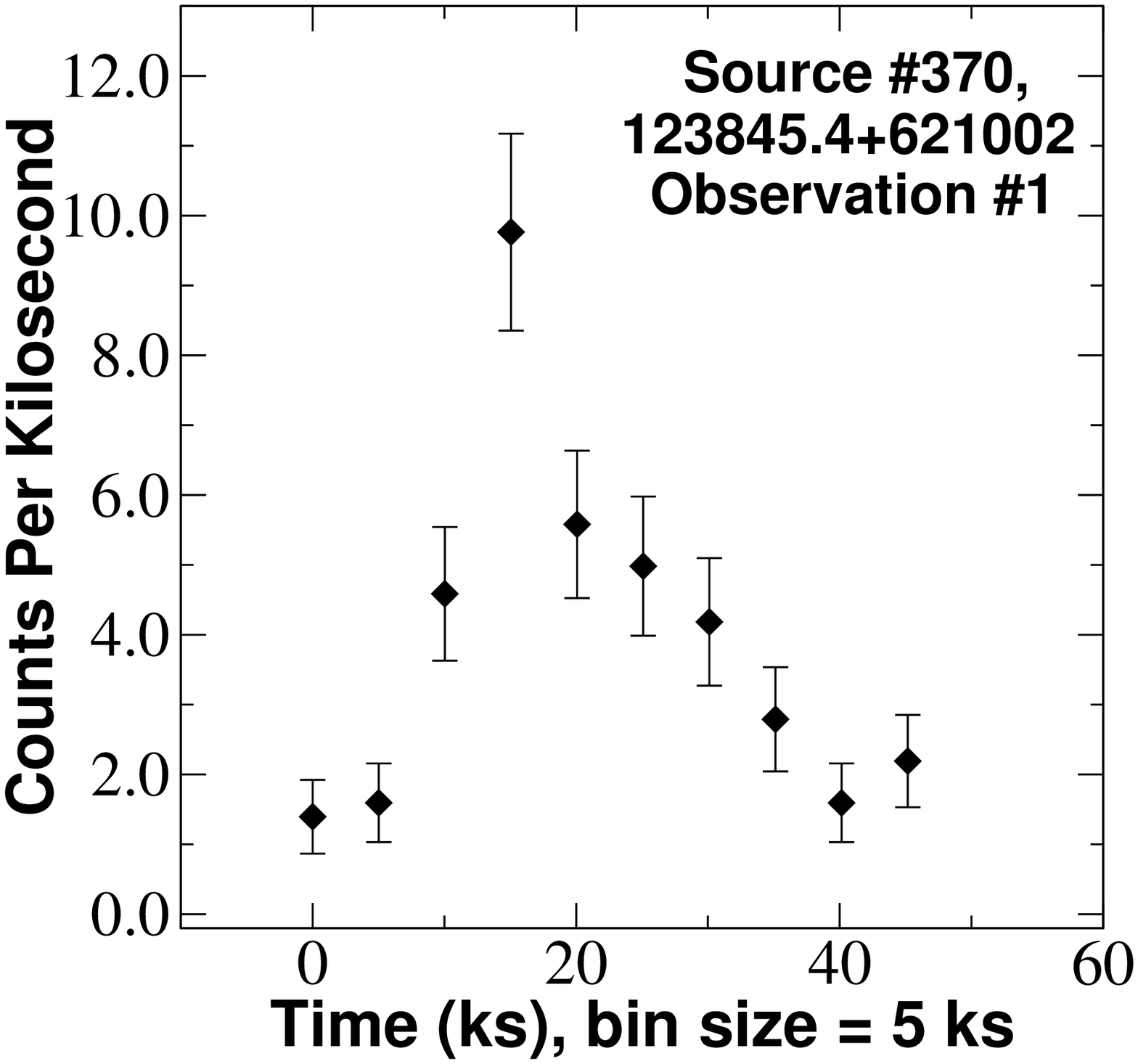}
  \includegraphics[width=0.3\textwidth]{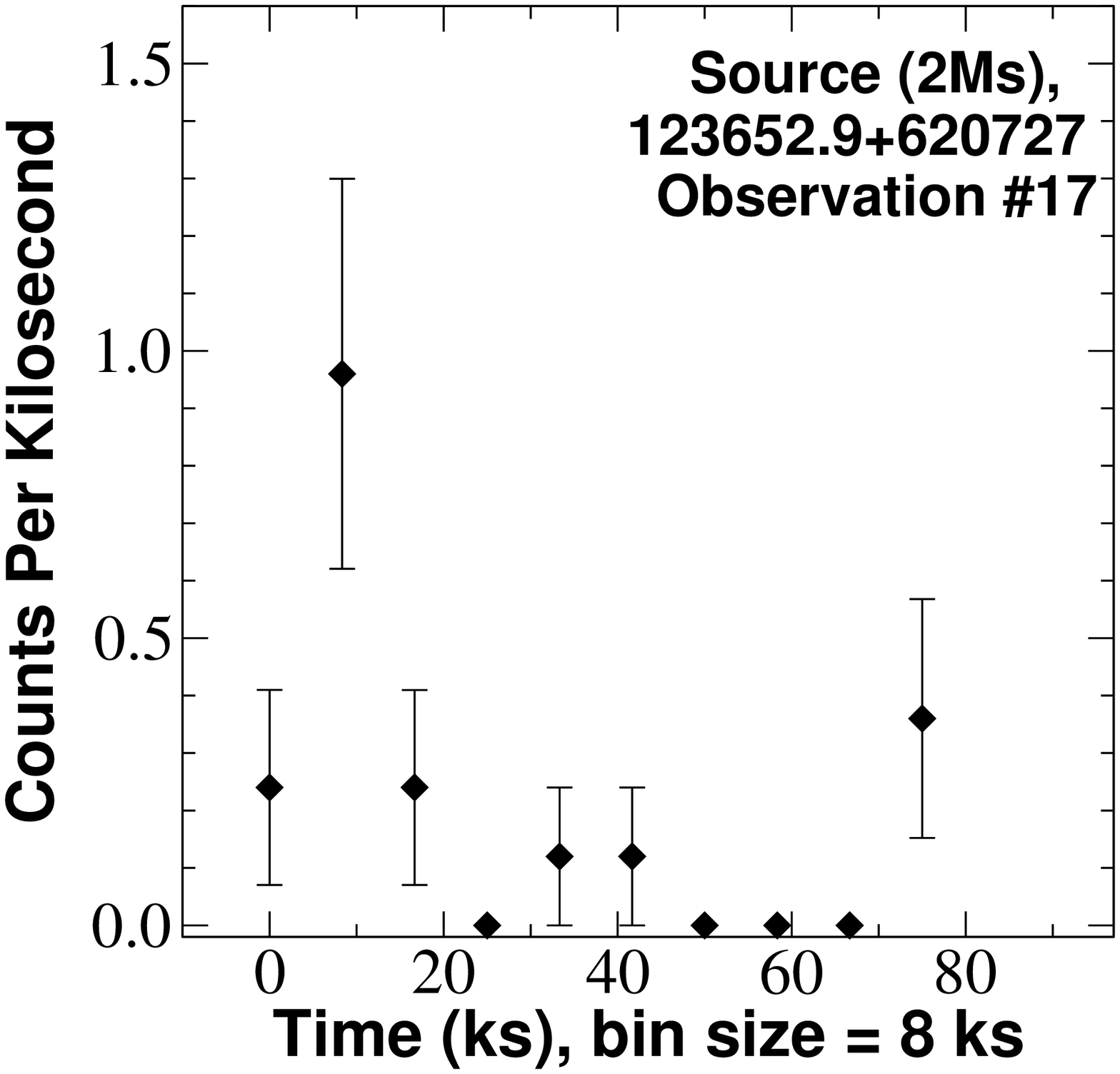}
\caption{Portions of the $Chandra$ ACIS lightcurves showing
statistically significant short-timescale flares from CDF-N stars.
The ordinate gives count rate in cts~ks$^{-1}$ and abscissa gives
time in ks. \label{xray_flares.fig}}
 \end{minipage}
\end{figure}

\clearpage
\newpage

\begin{figure}
 \begin{minipage}[t]{1.0\textwidth}
  \centering
    \includegraphics[width=0.45\textwidth]{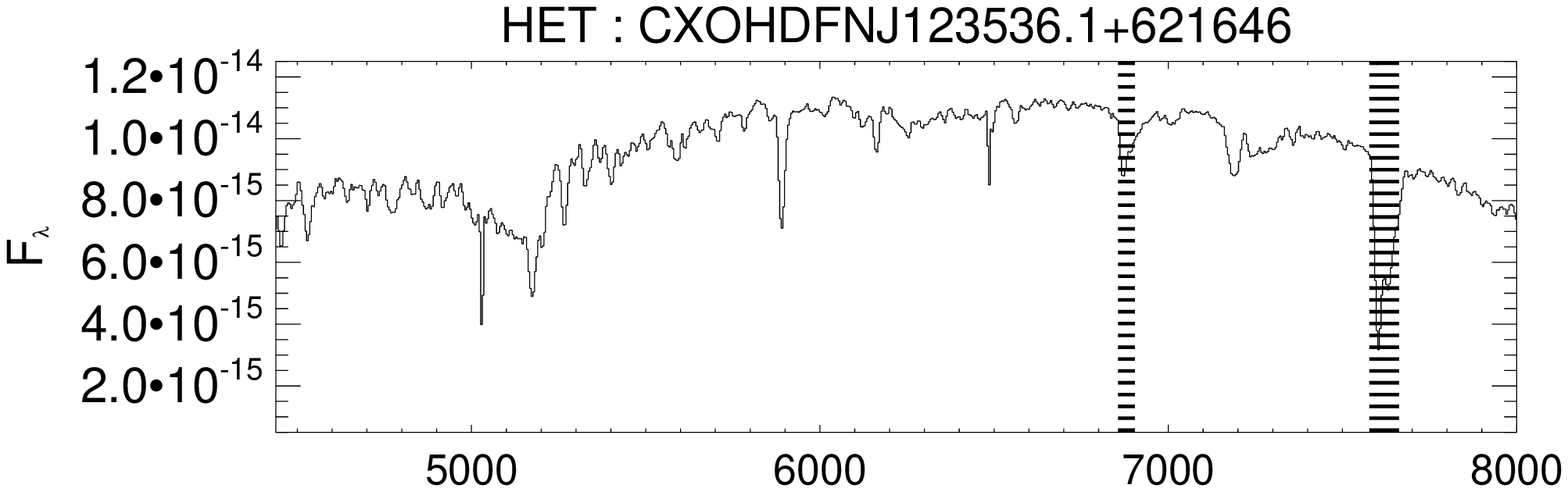}
    \includegraphics[width=0.45\textwidth]{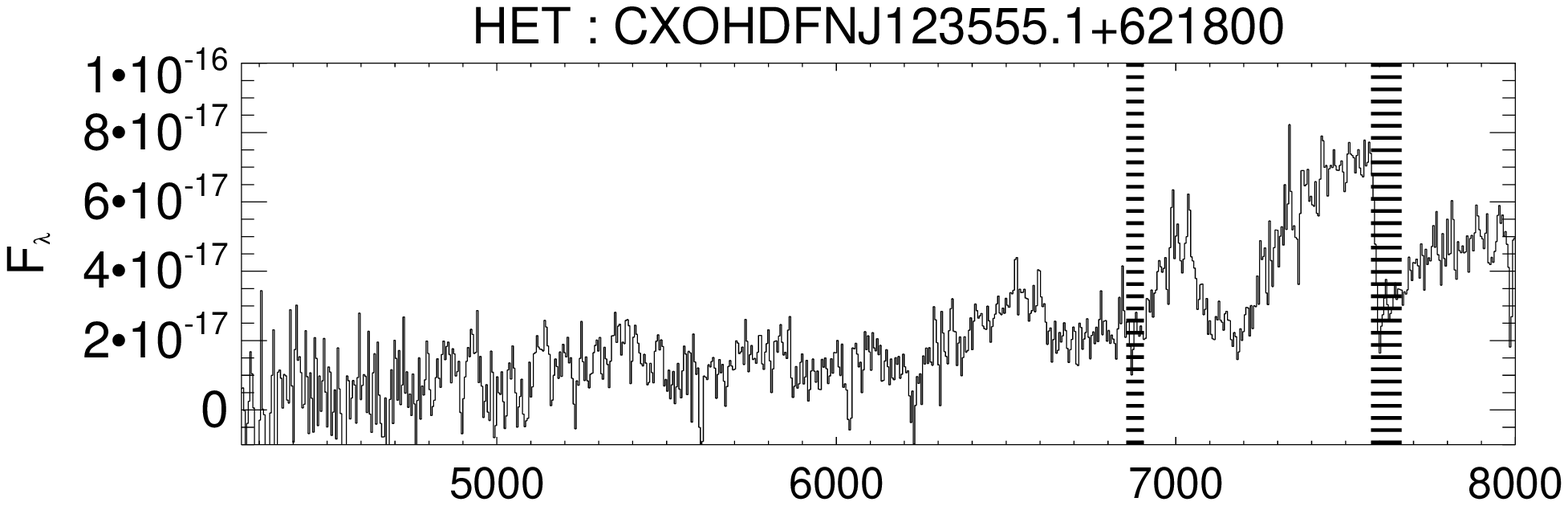}
 \end{minipage} \\ [0.2in]
 \begin{minipage}[t]{1.0\textwidth}
  \centering
    \includegraphics[width=0.45\textwidth]{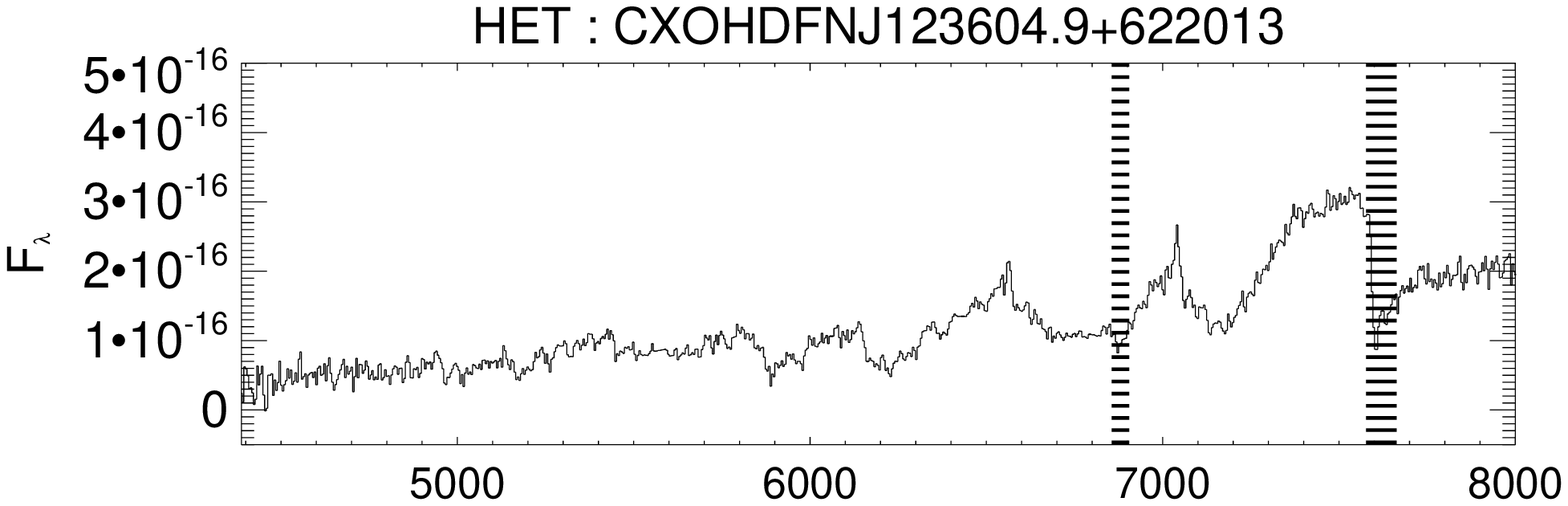}
    \includegraphics[width=0.45\textwidth]{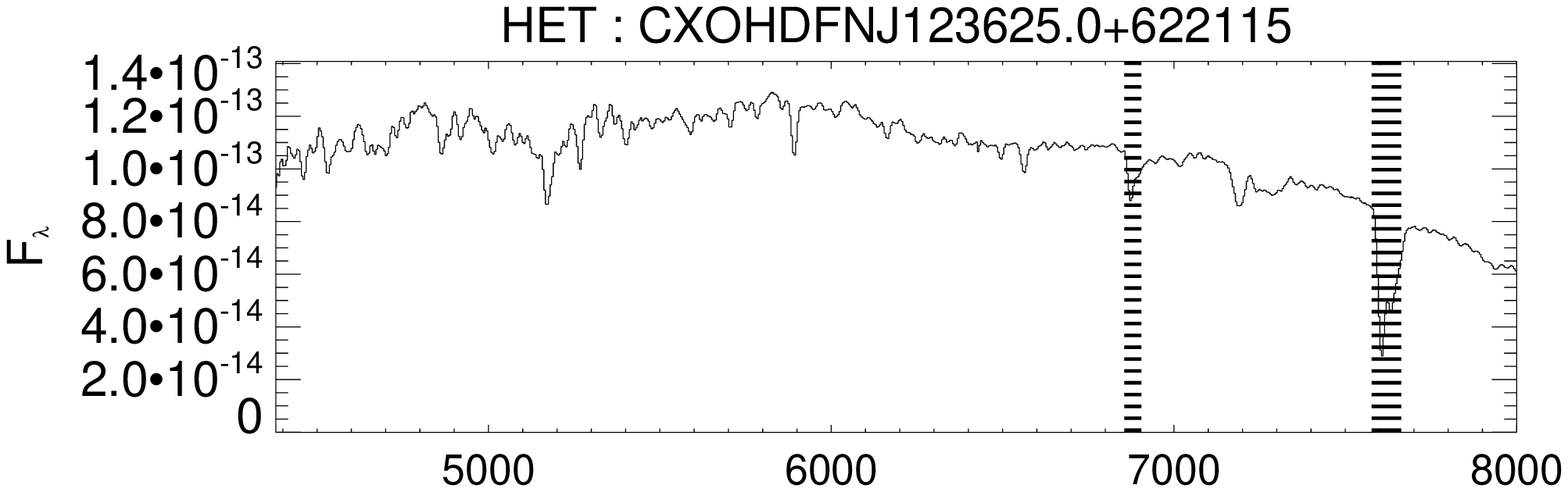}
 \end{minipage} \\ [0.2in]
 \begin{minipage}[t]{1.0\textwidth}
  \centering
    \includegraphics[width=0.45\textwidth]{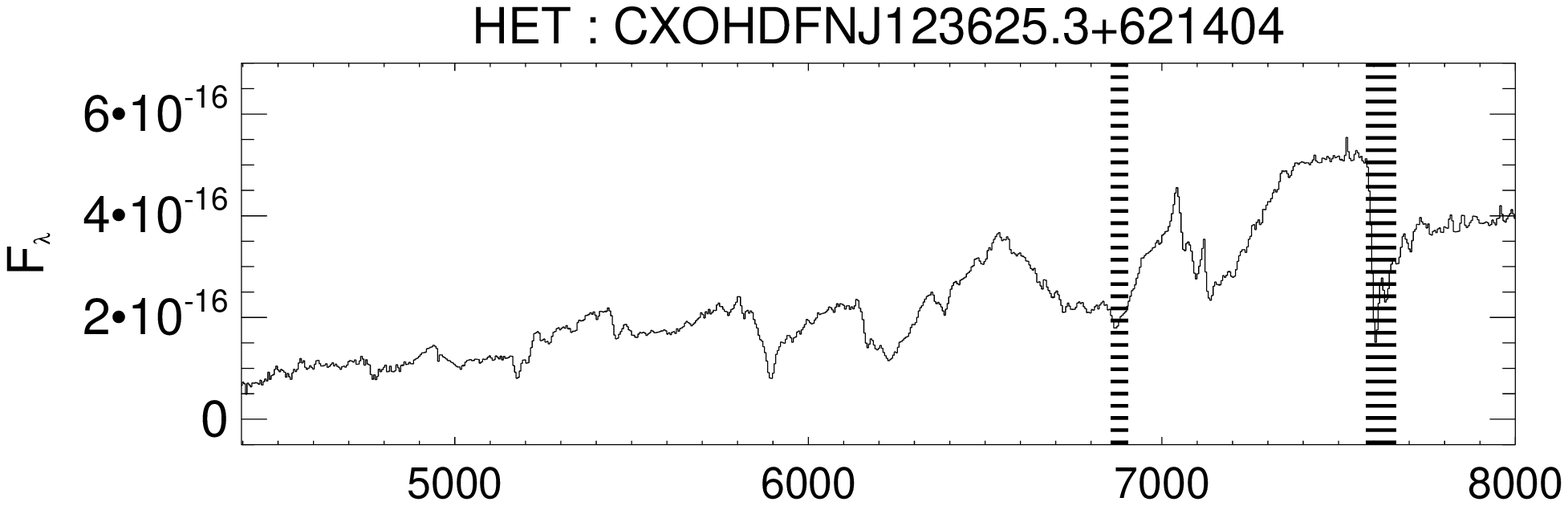}
    \includegraphics[width=0.45\textwidth]{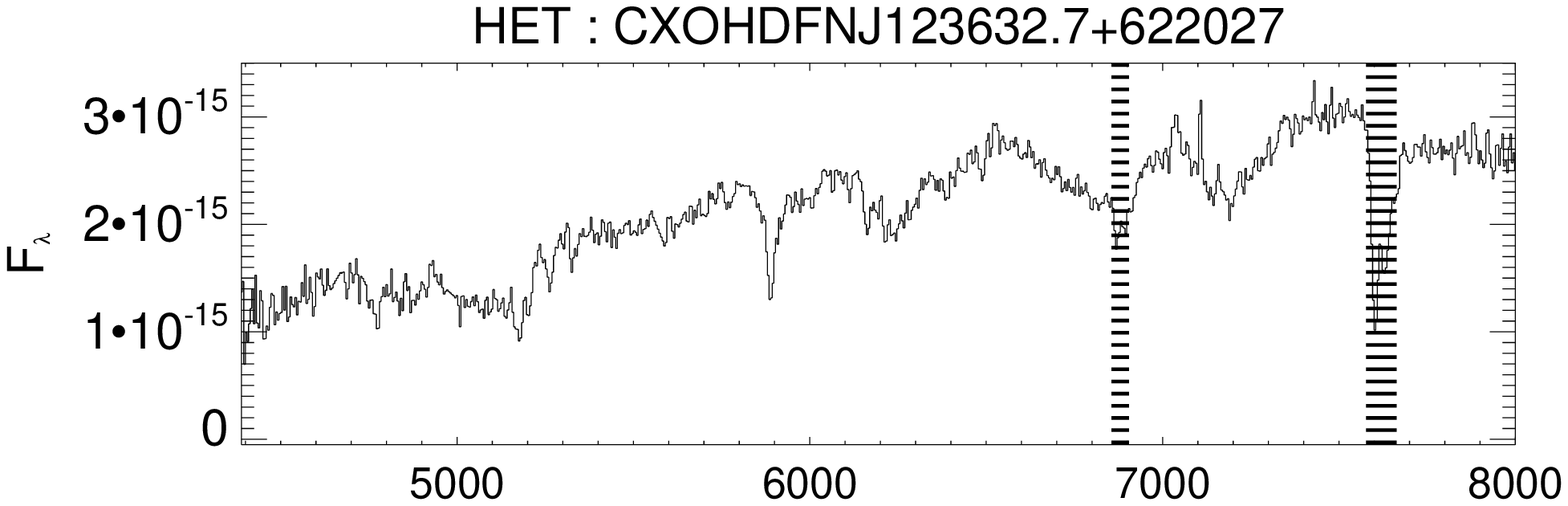}
 \end{minipage} \\ [0.2in]
 \begin{minipage}[t]{1.0\textwidth}
  \centering
    \includegraphics[width=0.45\textwidth]{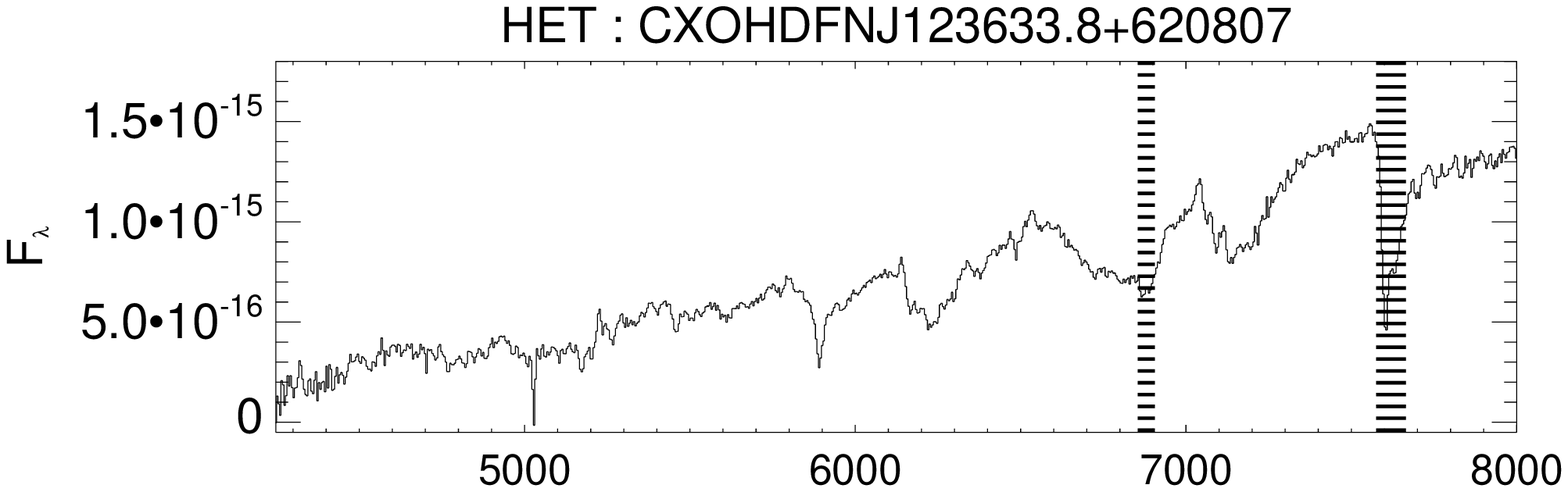}
    \includegraphics[width=0.45\textwidth]{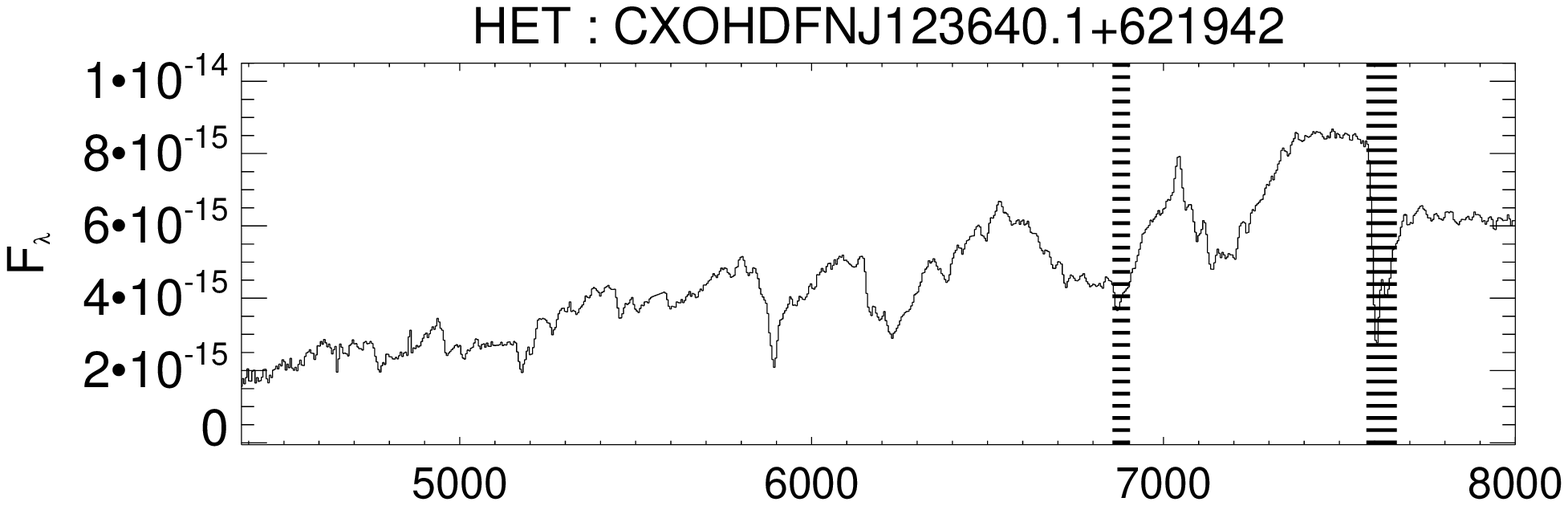}
 \end{minipage} \\ [0.2in]
 \begin{minipage}[t]{1.0\textwidth}
  \centering
    \includegraphics[width=0.45\textwidth]{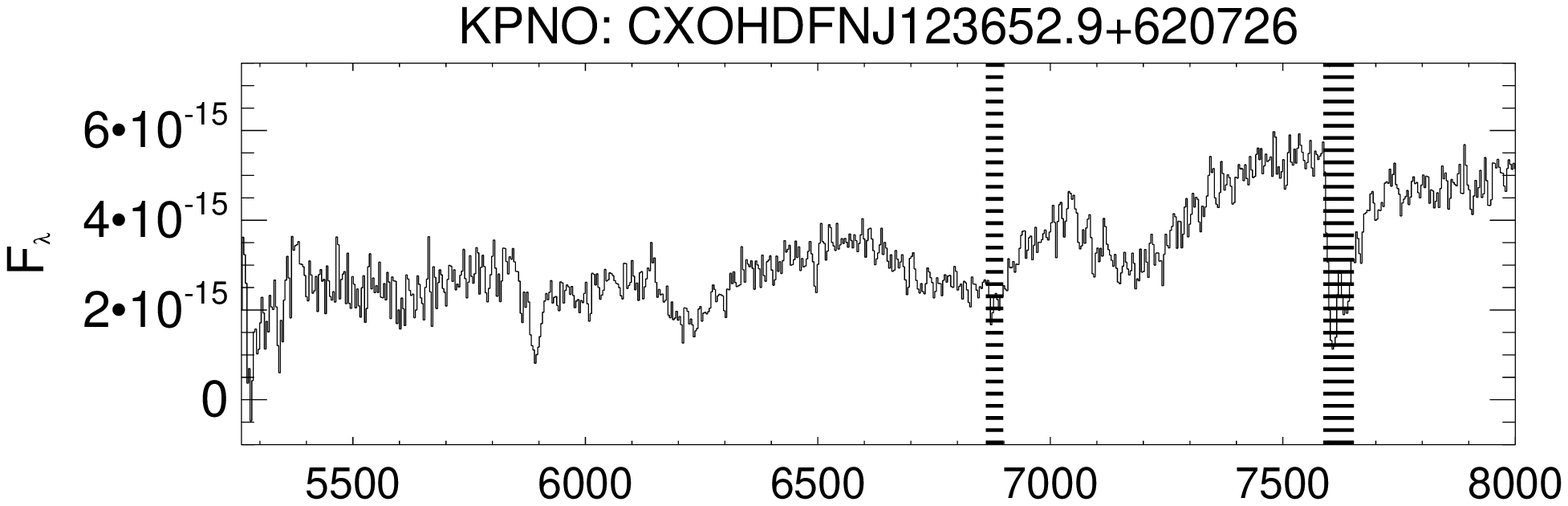}
    \includegraphics[width=0.45\textwidth]{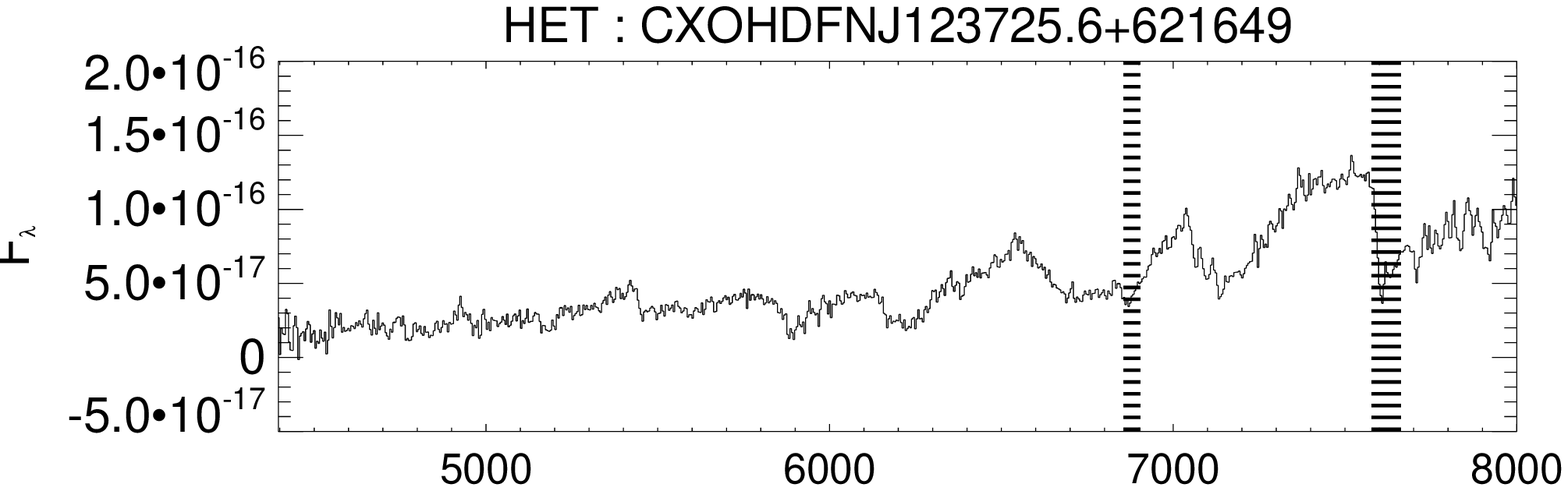}
 \end{minipage} \\ [0.2in]
 \begin{minipage}[t]{1.0\textwidth}
  \centering
    \includegraphics[width=0.45\textwidth]{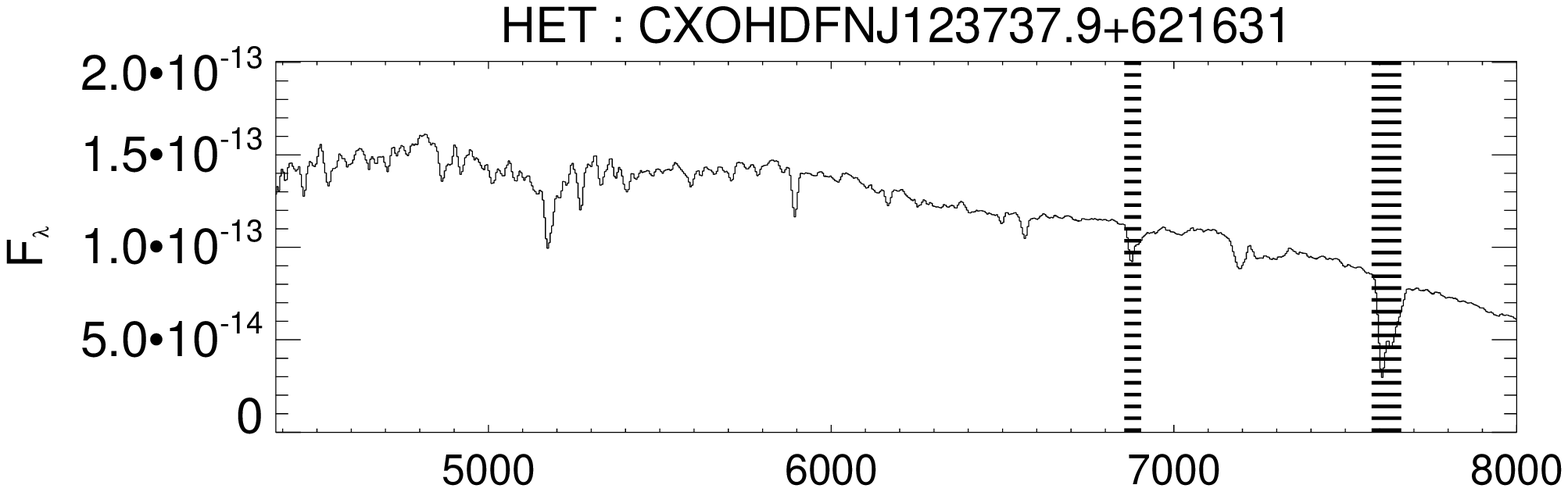}
    \includegraphics[width=0.45\textwidth]{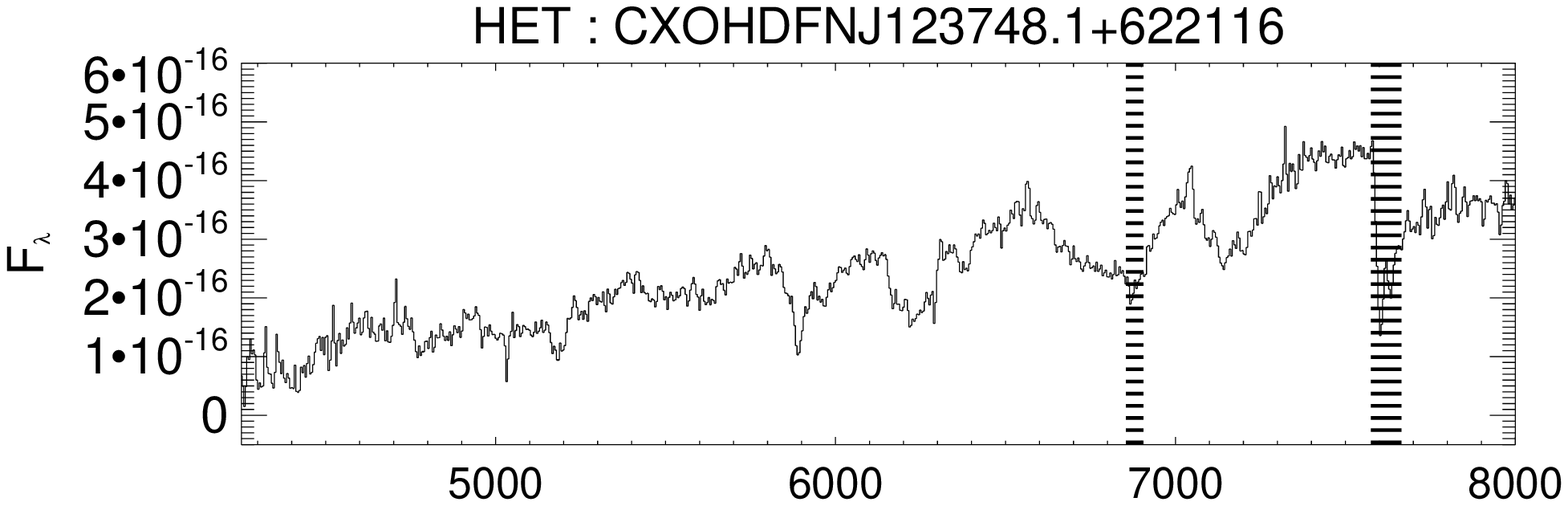}
 \end{minipage} \\ [0.2in]
 \begin{minipage}[t]{1.0\textwidth}
  \centering
    \includegraphics[width=0.45\textwidth]{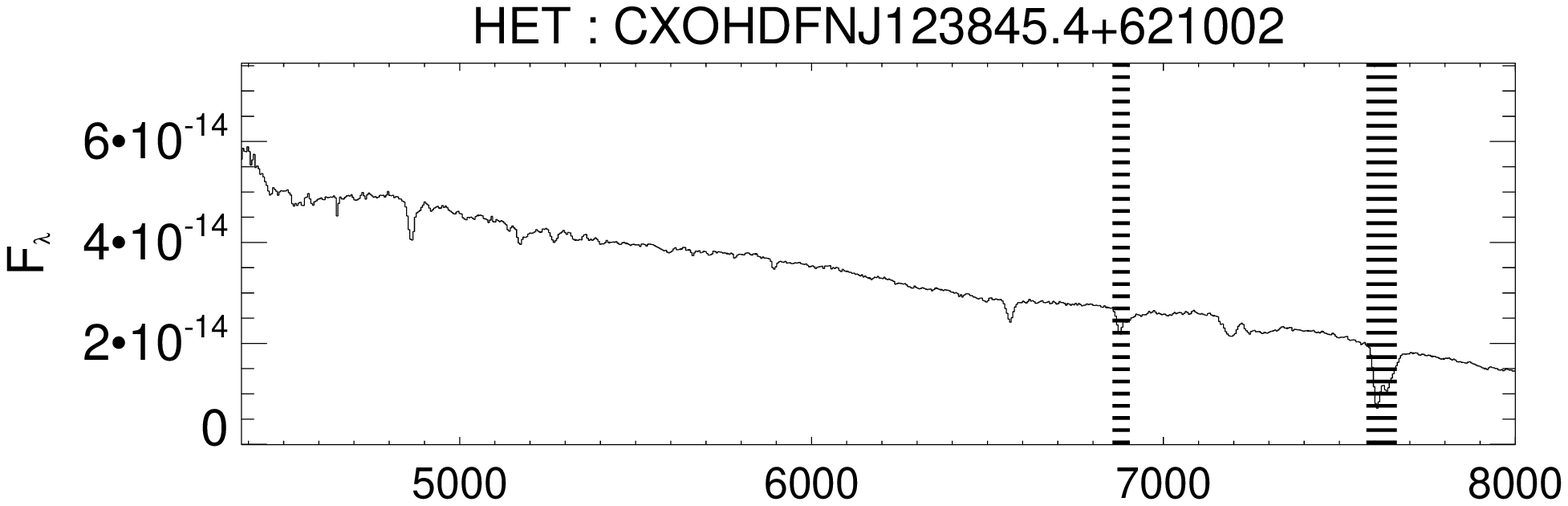}
\caption{Optical spectra of 13 CDF-N stars obtained with the
Hobby-Eberly Telescope 9m and Kitt Peak 4m telescopes.  The abscissa give
wavelength in Angstrom and the ordinates give flux densities in erg s$^{-1}$
cm$^{-2}$ $\AA^{-1}$.  Hatched areas denote spectral regions with telluric
contamination.  \label{opt_spec.fig}}
  \end{minipage}
\end{figure}

\clearpage
\newpage

\begin{figure}
\centering
  \includegraphics[width=0.32\textwidth]{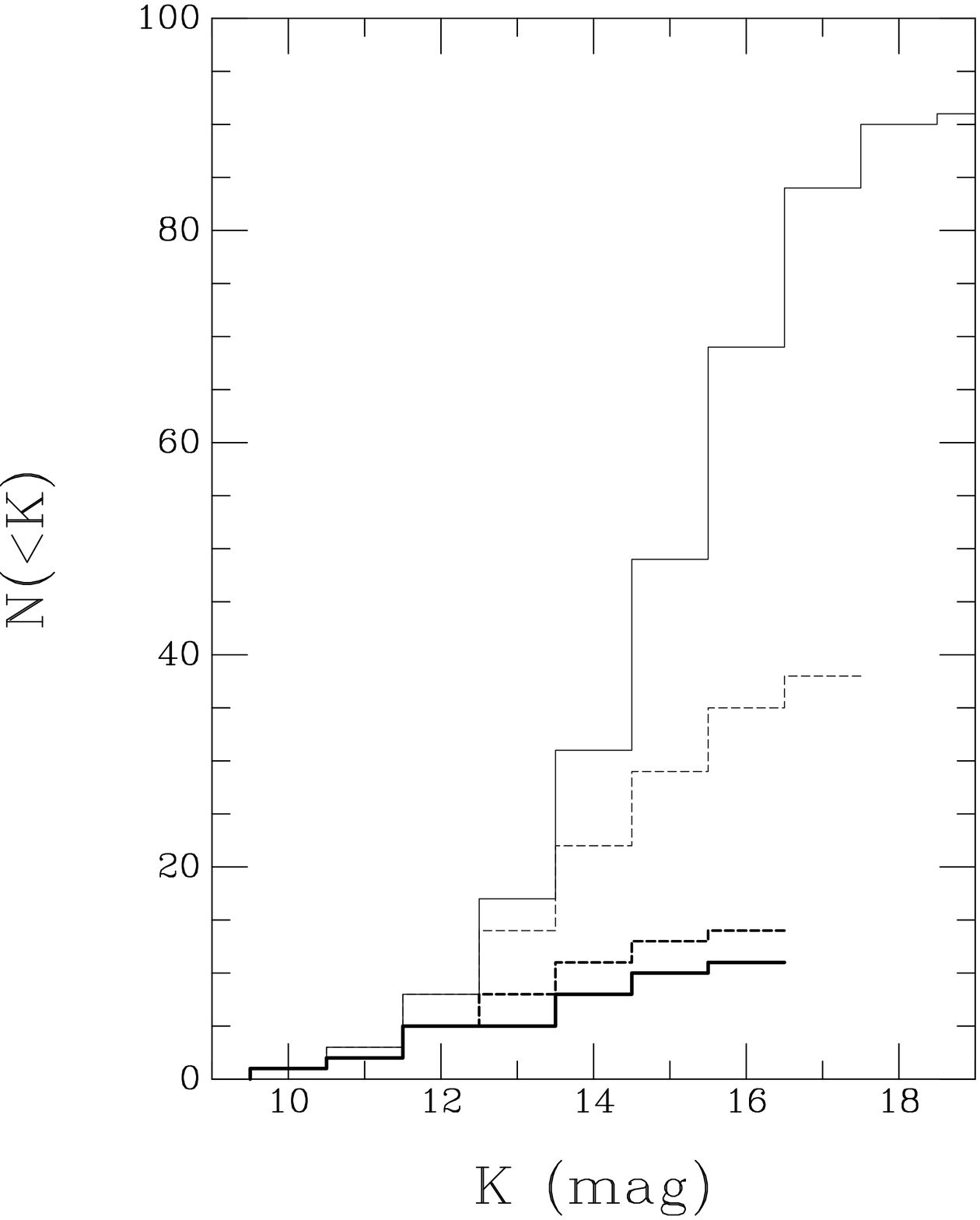}
  \includegraphics[width=0.32\textwidth]{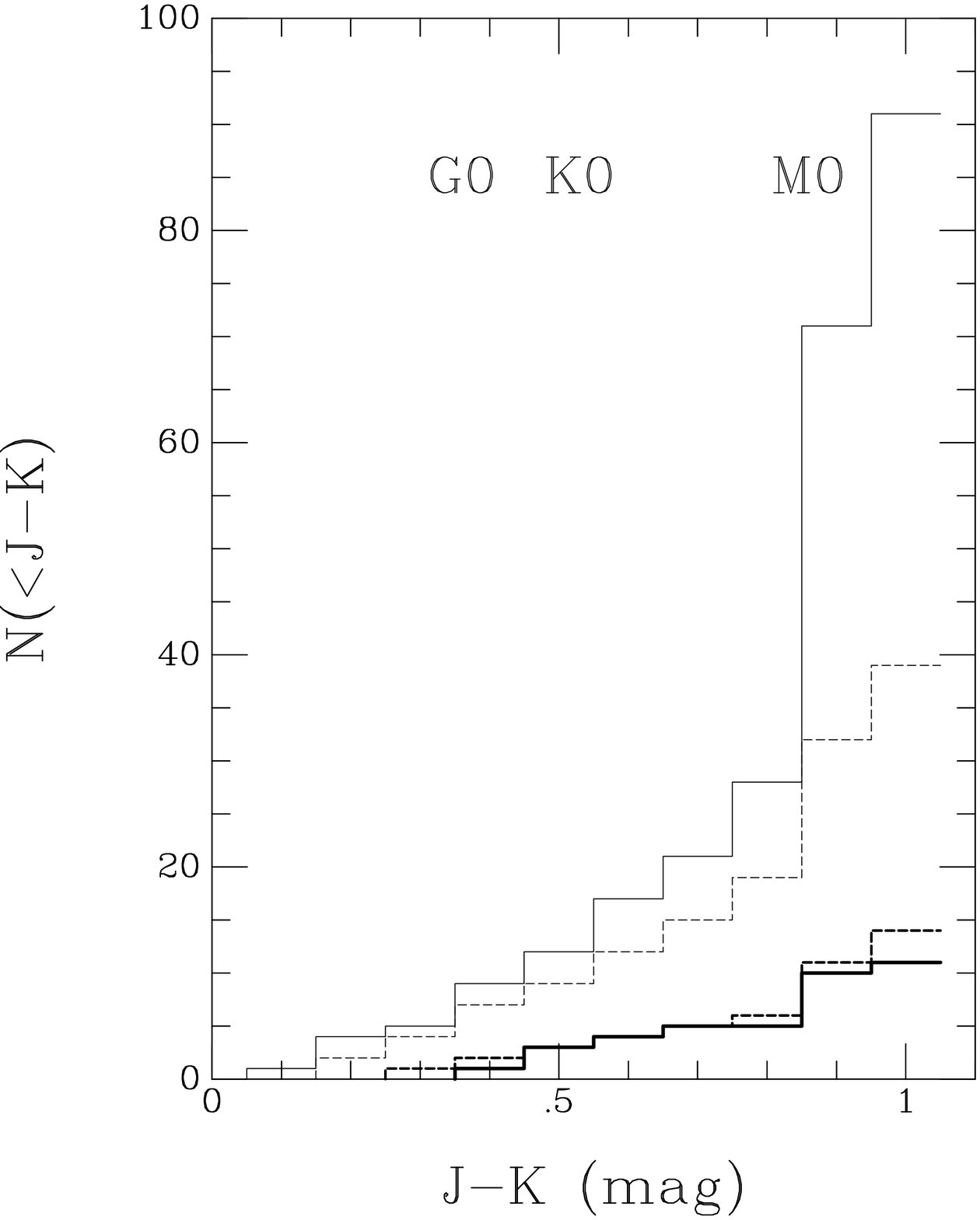}
  \includegraphics[width=0.32\textwidth]{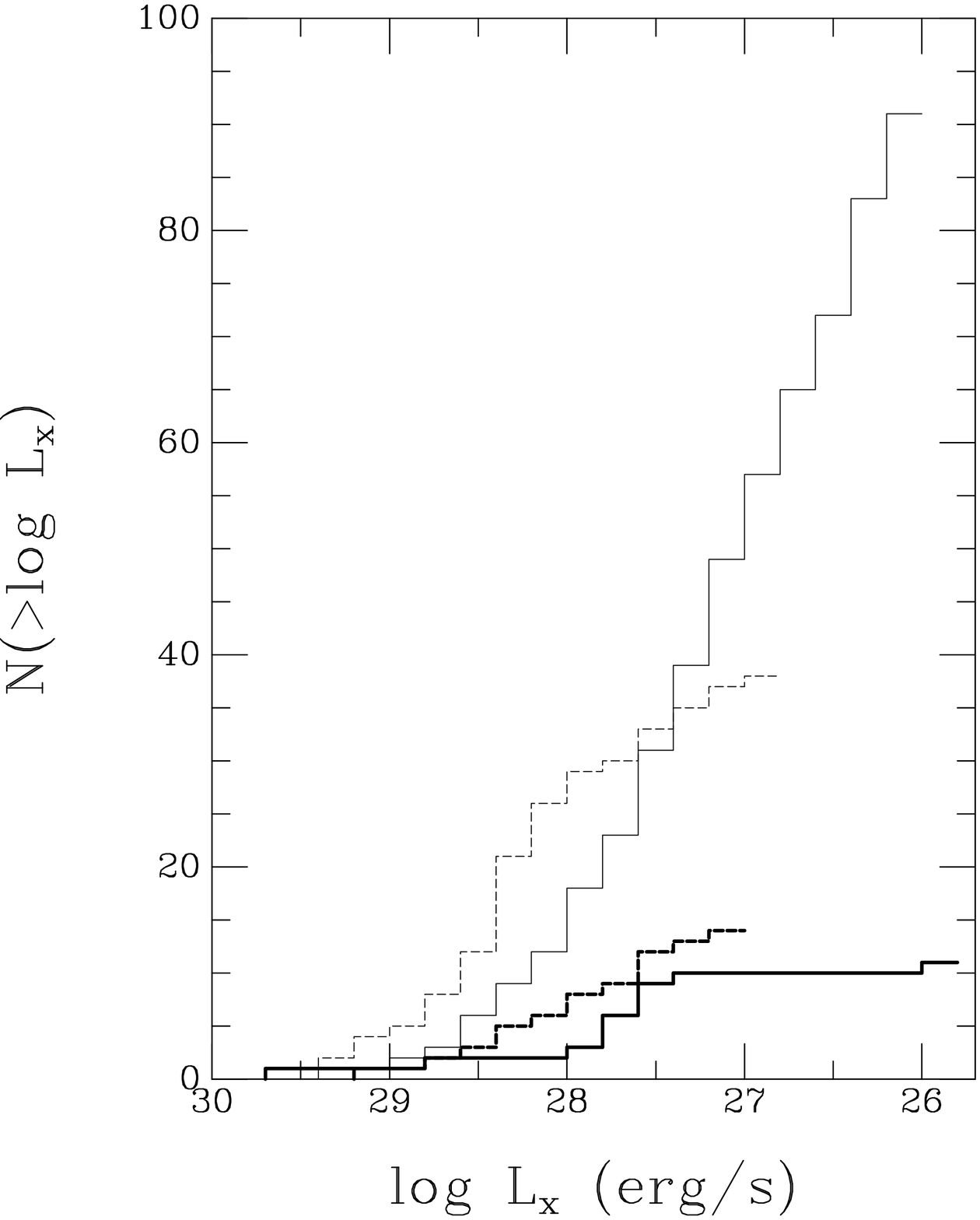}
  \caption{Comparison of cumulative distributions of three stellar
parameters for {\it XCOUNT} models of the CDF-N stellar
population: (left) $K$-band magnitude, (middle) $J-K$ color, and
(right) X-ray luminosity.  In each panel, histograms from top to
bottom are: total stellar population in the CDF-N field with
$V<22.5$ without X-ray selection (thin solid line), {\it XCOUNT}
model prediction with standard settings including X-ray selection
and no $t>1$ Gyr magnetic activity evolution (thin dashed line),
{\it XCOUNT} model with rapid X-ray decay with $\alpha \times
\beta = -2$ (thick dashed line), and the observed distributions
(thick solid line). \label{model.fig}}
\end{figure}

\clearpage
\newpage

\begin{deluxetable}{cc}
 \tablewidth{0pt}
 \tablecolumns{2}
 \tablecaption{CDF-N 1~Ms sky coverage \label{sens.tab}}

 \tablehead{
 \colhead{Count rate} &
 \colhead{Area} \\

 \colhead{(cts s$^{-1}$)} &
 \colhead{(deg$^{2}$)} }

\startdata
  $1.0\times10^{-5}$ &    0.0035 \\
  $1.2\times10^{-5}$ &    0.0212 \\
  $2.0\times10^{-5}$ &    0.0311 \\
  $4.0\times10^{-5}$ &    0.0386 \\
  $6.0\times10^{-5}$ &    0.0111 \\
  $8.0\times10^{-5}$ &    0.0038 \\
  $1.0\times10^{-4}$ &    0.0017 \\
  $1.0\times10^{-3}$ &    0.0112 \\
\enddata

\tablecomments{Count rates are in the $0.5-2$ keV soft band.}

\end{deluxetable}

\newpage

\begin{deluxetable}{rrrrrrrrr}
\rotate \tablecolumns{9}
\tabletypesize{\small} \tablewidth{0pt}
\tablecaption{X-ray positions and counts of CDF-N stars
\label{xray.tab}}

\tablehead{
 \colhead{} &
 \colhead{Coordinates (error)} &
 \colhead{Off-axis} &
 \multicolumn{3}{c}{Counts} &
 \colhead{Band} &
 \colhead{Ct Rate}&
 \colhead{$t_{eff}$}  \\ \cline{2-2} \cline{4-6}

 \colhead{B01 \#} &
 \colhead{$\alpha_{2000}$, $\delta_{2000}$ ($\arcsec$)} &
 \colhead{$\arcmin$} &
 \colhead{FB} &
 \colhead{SB} &
 \colhead{HB} &
 \colhead{Ratio} &
 \colhead{(ct Ms$^{-1}$)} &
 \colhead{(ks)}
}

\startdata
\multicolumn{9}{c}{\it Statistical 1~Ms Sample} \\

     12 & $123536.14+621646.5$ ~(1.2) &  8.6~~~~ & $127 \pm 20$ & $130 \pm 19$ & $<   30$    & $<0.24$          &  136~~~ &  948 \\
     36 & $123555.13+621800.3$ ~(0.9) &  7.2~~~~ & $ 53 \pm 14$ & $ 53 \pm 11$ & $<   21$    & $<0.41$          &   49~~~ & 1073 \\
     53 & $123604.90+622013.9$ ~(0.4) &  7.9~~~~ & $382 \pm 29$ & $365 \pm 29$ & $<   41$    & $<0.12$          &  418~~~ &  871 \\
    106 & $123625.01+622115.7$ ~(0.4) &  7.7~~~~ & $224 \pm 20$ & $214 \pm 17$ & $<   25$    & $<0.12$          &  204~~~ & 1046 \\
    108 & $123625.39+621404.8$ ~(0.3) &  2.4~~~~ & $519 \pm 25$ & $478 \pm 23$ & $ 33 \pm 8$ & $ 0.07 \pm 0.02$ &  317~~~ & 1509 \\
    128 & $123632.78+622027.2$ ~(0.4) &  6.7~~~~ & $396 \pm 30$ & $377 \pm 29$ & $<   29$    & $<0.08$          &  237~~~ & 1588 \\
    157 & $123640.12+621942.0$ ~(0.7) &  5.8~~~~ & $169 \pm 17$ & $166 \pm 15$ & $<   22$    & $<0.13$          &   95~~~ & 1741 \\
    302 & $123725.65+621649.0$ ~(0.7) &  5.5~~~~ & $ 97 \pm 14$ & $ 96 \pm 12$ & $<   20$    & $<0.22$          &   52~~~ & 1821 \\
    324 & $123737.99+621631.3$ ~(0.9) &  6.6~~~~ & $ 52 \pm 14$ & $ 49 \pm 12$ & $<   24$    & $<0.49$          &   29~~~ & 1671 \\
    341 & $123748.15+622126.8$ ~(1.5) & 10.4~~~~ & $195 \pm 23$ & $127 \pm 16$ & $ 79 \pm 17$& $ 0.64 \pm 0.16$ &  139~~~ &  917 \\
    370 & $123845.44+621002.5$ ~(0.7) & 14.5~~~~ & $218 \pm 18$ & $175 \pm 15$ & $ 34 \pm 9$ & $ 0.20 \pm 0.06$ & 4487~~~ &   39 \\
\\
\multicolumn{9}{c}{\it Additional 2 Ms stars}  \\
\nodata & $123633.81+620807.7$ ~(0.8) &  6.0~~~~ & $ 38 \pm 12$ & $ 36 \pm  9$ & $<   22$    & $<0.63$          &   20~~~ & 1785 \\
\nodata & $123652.95+620726.8$ ~(0.9) &  6.6~~~~ & $120 \pm 18$ & $110 \pm 16$ & $<   25$    & $<0.23$          &   62~~~ & 1763 \\
\enddata

\end{deluxetable}

\newpage

\begin{deluxetable}{rcrrrrrcc}
\centering
\tabletypesize{\small}
\tablewidth{0pt}
\tablecolumns{9}

\tablecaption{X-ray properties of CDF-N stars \label{xray_prop.tab}}
\tablehead{
\colhead{} &
\multicolumn{6}{c}{Spectrum} &&
\colhead{Var} \\
\cline{2-7} \cline{9-9}

 \colhead{B01~\#} &
 \colhead{$N_{\rm H}$} &
 \colhead{$kT_{1}$} &
 \colhead{$kT_{2}$} &
 \colhead{$\chi^{2}_{\nu}$~($\nu$)} &
 \colhead{$f_{\rm SB}$ } &
 \colhead{$f_{\rm FB}$ } &&
 \colhead{Num} \\

 \colhead{} &
 \colhead{($10^{21}$ cm$^{-2}$)} &
 \colhead{(keV)}&
 \colhead{(keV)} &
 \colhead{} &
 \multicolumn{2}{c}{($10^{-16}$ erg s$^{-1}$ cm$^{-2}$)} &&
 \colhead{Flares} }

\startdata
\multicolumn{7}{c}{\it Statistical 1~Ms sample} \\
 12     &  0.0 & $0.1$                  & $0.7$\tablenotemark{a} &  0.7 (10) &   6~~ &   6~~~~~~~ && 0 \\
 36     &  0.2 & $0.3$                  &   \nodata              &  1.8 ~~(4)&   3~~ &   3~~~~~~~ && 1 \\
 53     &  0.2 & $0.7$                  & $2.9$                  &  1.1 (13) &  19~~ &  24~~~~~~~ && 1 \\
106     &  0.2 & $0.5$                  &   \nodata              &  1.3 (17) &  10~~ &  10~~~~~~~ && 0 \\
108     &  0.2 & $0.6$                  & $1.5$                  &  1.2 (32) &  14~~ &  16~~~~~~~ && 3 \\
128     &  0.0 & $0.8$                  & $2.7$                  &  1.1 (11) &   9~~ &  11~~~~~~~ && 0 \\
157     &  0.0 & $0.2$\tablenotemark{b} & $0.8$                  &  0.8 (11) &   5~~ &   5~~~~~~~ && 1 \\
302     &  0.0 & $0.3$                  & $1.5$                  &  1.0 ~~(8)&   3~~ &   3~~~~~~~ && 1 \\
324     &  0.0 & \tablenotemark{c}      &   \nodata              &   \nodata &   1.7 &   1.8~~~~~ && 0 \\
341     &  0.2 &  \nodata               & $2.2$                  &  0.8 (27) &   6~~ &   9~~~~~~~ && 0 \\
370     &  0.0 & $0.1$                  & $1.4$                  &  1.3 (32) & 205~~ & 261~~~~~~~ && 1 \\
\multicolumn{7}{c}{\it Additional 2~Ms stars} \\
\nodata &  0.2 & 0.8                    &   \nodata              &   \nodata &   0.9 &   0.9~~~~~ && 0 \\
\nodata &  0.0 & $0.3$\tablenotemark{c} & $2.0$                  &  0.9 (11) &   4~~ &   4~~~~~~~ && 1 \\

\enddata

\tablenotetext{a}{An additional hard component with emission around 5 keV
may be present.}

\tablenotetext{b}{An additional ultra-soft component may be
present.}

\tablenotetext{c}{No good statistical fit was obtained for this
weak source. Fluxes were derived by assuming $kT = 0.8$ keV.}

\end{deluxetable}

\clearpage
\newpage

\begin{deluxetable}{rcccccccc}
\tabletypesize{\footnotesize} \tablewidth{0pt} \tablecolumns{9}
\tablecaption{Optical/IR properties of CDF-N stars
\label{phot.tab}}

\tablehead{ \colhead{~B01 \#} & \colhead{$\Delta^{\rm a}$ } &
\colhead{$B$} & \colhead{$V$} & \colhead{$R$} & \colhead{$I$} &
\colhead{$J$\tablenotemark{b}} & \colhead{$H$\tablenotemark{b}} &
\colhead{$K$\tablenotemark{b}} }

\startdata
\multicolumn{9}{c}{\it Statistical 1~Ms sample} \\
 12  &  0.6 & 17.59\tablenotemark{c}  & \nodata                  &  \nodata                 & \nodata                  &  12.60  &  12.06  &  11.92 \\
 36  &  0.5 & 22.50\tablenotemark{c}  &  21.25\tablenotemark{d}  &  20.15\tablenotemark{c}  &  17.79\tablenotemark{d}  &  16.23  &  15.81  &  15.22 \\
 53  &  0.5 & 20.82\tablenotemark{c}  &  19.58\tablenotemark{c}  &  18.19\tablenotemark{e}  &  \nodata                 &  15.41  &  14.66  &  14.48 \\
106  &  0.2 & \nodata                 &  14.53\tablenotemark{d}  &  13.27\tablenotemark{e}  &  13.56\tablenotemark{d}  &  11.35  &  10.88  &  10.79 \\
108  &  0.2 & 20.37\tablenotemark{c}  &  19.25\tablenotemark{d}  &  \nodata                 &  16.65\tablenotemark{d}  &  15.17  &  14.55  &  14.23 \\
128  &  0.1 & \nodata                 &  17.94\tablenotemark{d}  &  \nodata                 &  16.22\tablenotemark{d}  &  14.77  &  14.13  &  13.91 \\
157  &  0.5 & 18.80\tablenotemark{c}  &  16.57\tablenotemark{d}  &  \nodata                 &  15.07\tablenotemark{d}  &  12.87  &  12.24  &  12.01 \\
302  &  0.5 & 22.12\tablenotemark{c}  &  21.09\tablenotemark{d}  &  20.05\tablenotemark{c}  &  18.03\tablenotemark{d}  & \nodata & \nodata &  15.8~ \\
324  &  0.4 & \nodata                 &  14.02\tablenotemark{d}  &  13.12\tablenotemark{e}  &  13.08\tablenotemark{d}  &  10.70  &  10.26  &  10.19 \\
341  &  1.8 & 24.09\tablenotemark{c}  &  19.62\tablenotemark{d}  &  \nodata                 &  17.07\tablenotemark{d}  &  15.76  &  15.30  &  15.21 \\
370  &  0.3 & 14.17\tablenotemark{c}  &  \nodata                 &  \nodata                 & \nodata                  &  12.45  &  12.14  &  12.06 \\
\multicolumn{9}{c}{\it Additional 2~Ms stars} \\
\nodata& 0.2& \nodata                 &  16.81\tablenotemark{d}  &  \nodata                 &  14.86\tablenotemark{d}  &  13.19  &  12.55  &  12.28 \\
\nodata& 0.2& \nodata                 &  15.35\tablenotemark{d}  &  \nodata                 &  13.37\tablenotemark{d}  &  11.27  &  10.60  &  10.39 \\
\enddata

\tablenotetext{a} {Positional offsets in arcseconds of the ACIS
source with respect to the 2MASS Second Incremental Data Release
\citep{Cutri00}, except for source 302 where the optical source
position was measured by Barger et al. (2003). }

\tablenotetext{b} {JHK magnitudes are from 2MASS except for source
302 where we use the $HK^\prime$ measurement of \citet{Barger02}
with the approximate conversion $K = HK^\prime - 0.3$
\citep{Barger99}.}

\tablenotetext{c} {From \citet{Barger02}.}

\tablenotetext{d} {Measurement $V$ and $I$ plates of \citet{Wilson03}.}

\tablenotetext{e} {Measurement of $R$-band plate of Liu et al. (1999)
but note that for sources \#106 and \#324, Liu's measurements were
$R=12.77$ and $R=12.64$, respectively.}

\end{deluxetable}

\newpage

\begin{deluxetable}{rcrrc}
\tabletypesize{\footnotesize}
\tablewidth{0pt}
\tablecolumns{5}
\tablecaption{Derived properties of CDF-N stars \label{deriv.tab}}

\tablehead{ \colhead{B01 \#} & \colhead{SpTy} & \colhead{Dis} &
\colhead{$\log{ {f_{\rm X}}\over{f_K} } $} &
\colhead{$\log L_{SB}$} \\

\colhead{} &
\colhead{} &
\colhead{(pc)} &
\colhead{} &
\colhead{(erg s$^{-1}$)} \\
}

\startdata
\multicolumn{4}{c}{\it Statistical 1~Ms sample} \\
 12    & K4: &  320~~ & $-3.51$ &  27.7 \\
 36    & M4  &  360~~ & $-2.46$ &  27.5 \\
 53    & M5  &  160~~ & $-1.97$ &  27.8 \\
106    & K0  &  230~~ & $-3.72$ &  27.7 \\
108    & M4  &  230~~ & $-2.23$ &  27.9 \\
128    & M2  &  300~~ & $-2.53$ &  28.1 \\
157    & M5  &   50~~ & $-3.55$ &  26.1 \\
302    & M4  &  460~~ & $-2.28$ &  27.8 \\
324    & G5: &  220~~ & $-4.74$ &  26.8 \\
341    & M4  &  360~~ & $-2.20$ &  28.8 \\
370    & G8: &  510~~ & $-1.91$ &  29.8 \\
\multicolumn{4}{c}{\it Additional 2~Ms stars} \\
\nodata& M4  &   90~~ & $-4.20$ & 25.9 \\
\nodata& M2: &   75~~ & $-4.35$ & 26.2 \\
\enddata

\end{deluxetable}

\end{document}